\documentclass[superscriptaddress,longbibliography, reprint,amsmath,amssymb,aps,prx,floatfix]{revtex4-2}

\usepackage[margin=1.5cm]{geometry}

\usepackage{amssymb,amsmath,amsfonts,bm}

\usepackage{graphicx,amsmath,booktabs}%
\usepackage{color, array,amsmath,amsthm,amssymb,bm}
\usepackage{xcolor}
\usepackage[bookmarks=false]{hyperref}
\usepackage{subfiles}
\usepackage{braket}
\usepackage{hyperref}
\usepackage{CJKutf8}
\hypersetup{colorlinks=true,citecolor=blue,linkcolor=blue,urlcolor=blue,pdfstartview=FitH,bookmarksopen=true}

\makeatletter

\pdfpageheight\paperheight
\pdfpagewidth\paperwidth

\providecommand{\tabularnewline}{\\}

\begin{document}

\title{Hardware-Efficient Erasure Qubits With Superconducting Transmon Qutrits}

\author{Bao-Jie Liu}
\author{Ying-Ying Wang}

\affiliation{Department of Physics, University of Massachusetts-Amherst, Amherst, MA, USA}

\author{\begin{CJK}{UTF8}{gbsn}Yu-Xin Wang (王语馨)\end{CJK}}
\affiliation{Joint Center for Quantum Information and Computer Science,
University of Maryland, College Park, MD, 20742, USA}

\author{Manthan Badbaria}
\affiliation{Department of Physics, University of Massachusetts-Amherst, Amherst, MA, USA}

\author{Shruti Puri}
\affiliation{Yale Quantum Institute and Department of Applied Physics,  ale University, New Haven, CT 06511}

\author{Chen Wang}
\email[Email: ]{wangc@umass.edu}

\affiliation{Department of Physics, University of Massachusetts-Amherst, Amherst, MA, USA}

\date{\today}%
\begin{abstract}
Quantum error correction using erasure qubits offers higher fault-tolerant thresholds and improved scaling by converting dominant physical errors into detectable erasures. %~\bj{Erasure qubits have been experimentally demonstrated in circuit QED to provide long memory times in both dual-rail transmon and cavity-based implementations.} 
In superconducting circuits, erasure qubits can be constructed using the dual-rail approach, which, however, requires additional qubit-count overhead and tailored coupling elements. 
Here, we demonstrate a hardware-efficient scheme that operates transmon qutrits as erasure qubits, which is compatible with standard superconducting circuit-QED hardware. The logical states $\ket{0_\text{L}}$ and $\ket{1_\text{L}}$ are represented by the ground and second excited states, while the dominant relaxation errors %decay from $\ket{1_\text{L}}$ to the first excited state %produces a detectable erasure. Erasure detection is performed 
can be detected via an ancilla qubit using a microwave-activated two-qutrit SWAP gate. We demonstrate a logical qubit $T_1$ lifetime exceeding $500\,\mu\mathrm{s}$, post-selected with repeated mid-circuit erasure detection, which is ten times longer than the $T_1$ time of the transmon physical qubit. Coherence times beyond $300\,\mu\mathrm{s}$ are achieved using dynamical decoupling. 
Single-qubit gate operations reach average Clifford gate infidelity on the order of $10^{-4}$. % of $7.6\times10^{-4}$. 
We further demonstrate dual-purposing an ancilla qubit for both erasure detection and parity checking, showing heralded generation of Bell states between erasure qubits. 
These results suggest that mainstream architectures of transmon qubit arrays may already be capable of implementing erasure-based QEC strategies for hardware-efficient fault-tolerant quantum computing. 

\end{abstract}
\maketitle
\def\thefootnote{*}

\begin{figure}[t]
\centering
\includegraphics[width=84mm]{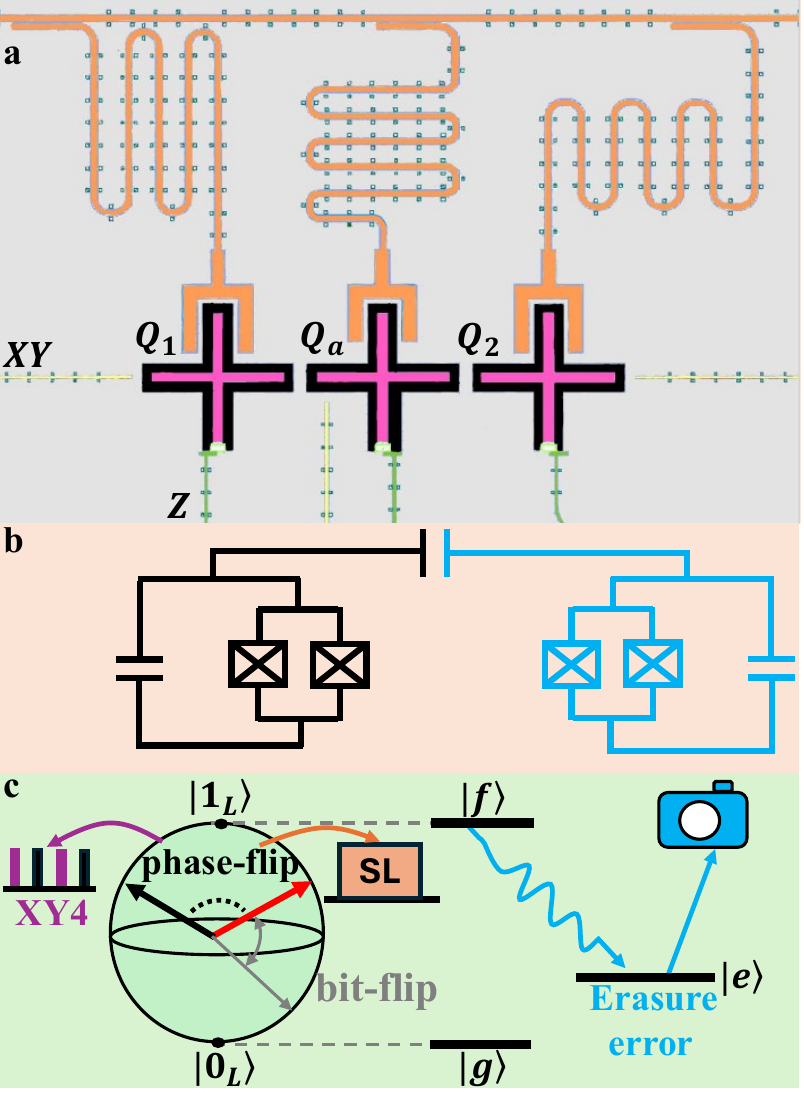}
\caption{ \textbf{Device schematic and $g$–$f$ erasure qubit scheme.} \textbf{a}, False-colored device layout of the three flux-tunable transmon qubits. \textbf{b}, Simplified effective
circuit of data qubit ($Q_1$) (black) coupled to an ancilla qubit ($Q_a$) (blue) used for erasure detection. \textbf{c}, Erasure qubit  is  encoded in the subspace of transmon qutrit level as $|0_\text{L}\rangle\equiv|g\rangle$ and $|1_\text{L}\rangle\equiv|f\rangle$. Decay from $|1_\text{L}\rangle$ produces a detectable leakage to $|e\rangle$, which is extracted and then detected via a microwave pulse on the ancilla resonant. % only when the data qubit is in $|e\rangle$. 
Phase-flip errors of the $g$–$f$ qubit can be suppressed by using XY4 or spin-locking sequences.} 
\label{fig1}
\end{figure}

\begin{figure*}[tbp]
\centering
\includegraphics[width=180mm]{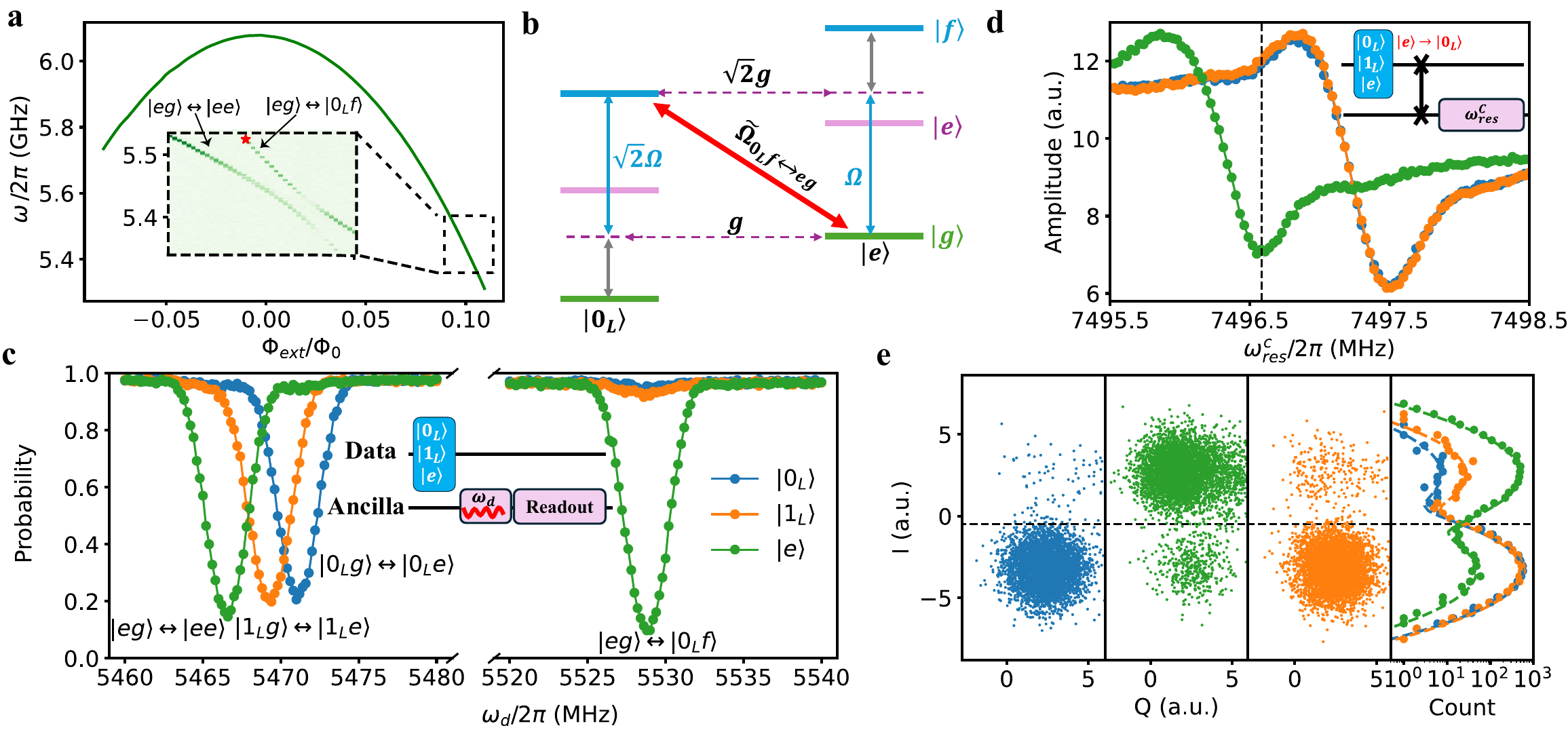}
\caption{\textbf{Ancilla-assisted erasure detection}. \textbf{a}, Frequency spectrum of the ancilla qubit as a function of its flux bias $\Phi_{\mathrm{ext}}$ (in units of $\Phi_{0}$). The inset shows an avoided level crossing between the states $|ee\rangle$ and $|0_\text{L}f\rangle$. \textbf{b}, Energy level structure of microwave-activated erasure detection. An erasure SWAP gate between $|eg\rangle$ and $|0_\text{L}f\rangle$ is realized by four-wave-mixing resonant parametric interactions with the effective coupling strength $\tilde{\Omega}_{0_\text{L}f \leftrightarrow e g}$, which is related to microwave drive strength $\Omega$ and coupling strength $g$ of data and ancilla qubit. \textbf{c}, Spectroscopy of the ancilla qubit, transitions of the ancilla are shown for the data qubit prepared in \(|0_\text{L}\rangle\) (blue), \(|1_\text{L}\rangle\) (orange), and \(|e\rangle\) (green). The observed frequency shifts of the ancilla's  \(|g\rangle \leftrightarrow |e\rangle\) transition correspond to the different states of the data qubit, arising from the static \(ZZ\) coupling between the data and ancilla qubit. The \(|g\rangle \leftrightarrow |f\rangle\) two-photon transition happens when the data qubit is in the \(|e\rangle\) state, which is used for erasure detection. \textbf{d}, We perform spectroscopy of the ancilla resonator with an erasure SWAP gate, after initializing the $g$–$f$ data qubit in $\ket{0_\text{L}}$, $\ket{1_\text{L}}$, or $\ket{e}$. The black dashed line indicates the readout drive frequency used for erasure detection. \textbf{e}, Distribution of the I--Q cloud obtained from ancilla readout measurements for the $\ket{0_\text{L}}$, $\ket{1_\text{L}}$, and $\ket{e}$ states of the data qubit. The right panel shows the histograms of the integrated I value, with dashed lines represent fits to a double Gaussian model. The horizontal black dashed line marks the qubit-state threshold.}
\label{fig2}
\end{figure*}

\section{Introduction}
Quantum error correction (QEC) is essential for realizing practical quantum computation, as it allows exponential suppression of errors arising from %protects fragile quantum states from
decoherence and operational noise 
%with increasing qubit count
~\cite{Shor1996,CalderbankShor1996,KnillLaflamme1997,Terhal2015,FowlerMariantoniMartinisCleland2012}.
%By keeping the error rates of physical qubits below a critical threshold~\cite{Terhal2015,FowlerMariantoniMartinisCleland2012}, QEC can exponentially suppress logical errors by redundantly encoding multiple physical qubits into a single data qubit. 
Recent experiments have demonstrated QEC operations at or beyond the break-even threshold~\cite{Sivak2023,Ni2023,GoogleQAI2025,QuditQEC2025,Li2025,Sun2025A,Shirol2025,Ni2025A}, but achieving algorithmically relevant logical error rates by scaling up existing hardware remains prohibitively demanding on both the quality and quantity of physical qubits. To obtain better scaling of error rates in QEC implementations, a promising approach is %to exploit a specific noise characteristics of each quantum platform. One particularly promising approach is the 
to leverage the favorable noise characteristics of erasure qubits~\cite{BennettDiVincenzoSmolin1997,GrasslBethPellizzari1997}. In an erasure qubit, the logical $\ket{0}$ and $\ket{1}$ are represented in such a way that Pauli errors are strongly suppressed, whereas the dominant decoherence errors manifest as detectable leakage from the computational space known as erasures. %In an erasure qubit, certain physical error processes (such as leakage out of the computational subspace or photon loss) are designed to produce a detectable flag indicating that an error has occurred and, crucially, which qubit was affected. 
QEC codes can tolerate erasures at significantly higher error rates, leading to more resource-efficient encodings and higher fault-tolerance thresholds compared to conventional schemes~\cite{kubica2023erasure,gu2025fault,gu2024optimizing,chang2024surface,Jacoby2025}. Recent experiments have realized erasure qubits in several physical platforms including neutral atoms~\cite{Wu2022Erasure,sahay2023high,scholl2023erasure,ma2023high,zhang2025leveraging}, trapped ions~\cite{Kang2023,quinn2024high}, and superconducting circuits~\cite{Campbell2020,teoh2023dual,chou2024superconducting,mehta2025bias,Koottandavida2024,de2025mid,levine2024demonstrating,huang2025logical,wills2025error}.

%neutral atoms and trapped ions with leakage detection based encodings~\cite{Wu2022Erasure,sahay2023high,scholl2023erasure,ma2023high,zhang2025leveraging,Kang2023,quinn2024high}, and superconducting cavities or transmons by using dual-rail scheme~\cite{teoh2023dual,Koottandavida2024,chou2024superconducting,de2025mid,mehta2025bias,levine2024demonstrating,huang2025logical,wills2025error}.

%To further improve fault tolerance, specialized strategies exploit the specific noise characteristics of each quantum platform. One promising strategy involves erasure qubits~\cite{BennettDiVincenzoSmolin1997,GrasslBethPellizzari1997}, in which the location of an error is known, enabling more efficient encoding and higher fault-tolerant thresholds than conventional encoding schemes~\cite{kubica2023erasure,gu2025fault,gu2024optimizing,chang2024surface,Jacoby2025}. Recent experiments have realized erasure qubits in neutral atoms and trapped ions with leakage detection based encodings~\cite{Wu2022Erasure,sahay2023high,scholl2023erasure,ma2023high,zhang2025leveraging,Kang2023,quinn2024high}, and superconducting cavities or transmons by using dual-rail scheme~\cite{teoh2023dual,Koottandavida2024,chou2024superconducting,de2025mid,mehta2025bias,levine2024demonstrating,huang2025logical,wills2025error}. 

In superconducting circuits, current implementations of erasure qubits are based on dual-rail encoding, which uses the $\ket{01}$ and $\ket{10}$ states of two physical qubits or cavities to represent a logical qubit.  The requirement of two storage modes per erasure qubit plus ancillary circuits for erasure detection leads to a sizable hardware overhead that partially compromises the scalability advantages of erasure-based QEC~\cite{chang2024surface}. The need for tailored frequency placements or specialized coupling schemes further increases the circuit complexity.  
%The requirement of multiple physical modes per erasure qubit increases circuit complexity, control wiring, and chip-area consumption. 
In contrast, an alternative proposal dubbed the $g$–$f$ erasure qubit~\cite{kubica2023erasure} exploits the multilevel structure and favorable noise properties already present in individual physical qubits such as the transmons. By representing logical information in the $\ket{g}$ (ground) and $\ket{f}$ (second excited) states of a three-level transmon qutrit, decay errors predominantly lead to leakages to the $\ket{e}$ state that are in principle detectable, while dephasing errors can be effectively suppressed by dynamic decoupling.  
By eliminating the hardware overhead %and the tailored design 
necessary for dual-rail encoding, the $g$–$f$ qubit may enable more efficient QEC strategies that are more readily adaptable on prevailing superconducting quantum processors. 
%By exploiting higher excited states already accessible in standard transmon devices, $g$–$f$ erasure qubits support erasure detection with minimal additional hardware, making them potentially suited for near-term, scalable implementations of erasure-based QEC. 

However, to operate regular transmons as $g$–$f$ erasure qubits, a core requirement remains to be fulfilled, namely high-fidelity mid-circuit erasure detection with minimal back-action on logical states. Direct dispersive readout of the $\ket{e}$ state without dephasing the $\ket{g}-\ket{f}$ superposition would be ideal but is generally impractical for the transmon. On the other hand, recent developments of transmon two-qutrit gates~\cite{Roy2023,Luo2023,Blok2021,nguyen2024empowering,goss2022high,Fischer2023} offer a practical route towards erasure detection using an ancilla transmon, although challenges to protect the computational states due to the small transmon anharmonicity must be carefully addressed. 

In this work, we demonstrate operation of $g$–$f$ erasure qubits based on the superconducting transmon qutrit. 
We realize erasure detection via an erasure-syndrome SWAP operation, driven by a microwave-activated four-wave mixing process between an ancilla and the data qubit. 
With repeated erasure detection and dynamic decoupling techniques, we achieve logical $T_1$ time exceeding $500\,\mu\mathrm{s}$ and $T_2$ time exceeding $300\,\mu\mathrm{s}$, %about an order of magnitude longer than the coherence times of the bare transmon.  
compared to $T_1^{ge}\approx55$ $\mu$s and $T_{2E}^{ge}\approx75$ $\mu$s of the regular transmon qubit. The logical gates on the $g$–$f$ qubit are implemented via a two-photon transition, showing an average error per gate of $4\times10^{-4}$ in randomized benchmarking. 
Furthermore, by reusing the erasure-detection ancilla for two-qubit parity measurements, we %demonstrate both erasure detection and parity checking, showing 
show heralded generation of Bell states between two $g$–$f$ erasure qubits. These results suggest that the $g$–$f$ qubit may provide a promising route toward hardware-efficient quantum error correction by embedding both error-resilient encoding and error-detection capabilities within a minimal physical footprint~\cite{Margaret2026}. %Looking forward, this architecture could enable scalable designs where error monitoring and logical operations are naturally integrated, reducing overhead while maintaining high performance~\cite{Margaret2026}. \cw{Need a more insightful and less generic last sentence.}

\section{Microwave-based Erasure Detection}
We first describe erasure detection of the $g$–$f$ qubit. The goal is to perform a gate between the data qubit and an ancilla such that the ancilla is excited if and only if the data qubit occupies the $|e\rangle$ state, followed by an ancilla readout. As shown in Fig.~\ref{fig1}, our study is carried out on a planar three-qubit device with a common X-shaped transmon design~\cite{Chen2014}, fairly standard set of circuit Hamiltonian parameters, and modest coherence times (see Supplementary Information). %We note that the device has a fairly standard set of circuit parameters and was not designed specifically for erasure qubits. 
We utilize two of the three transmons ($Q_{1}$ and $Q_{a}$) to show experiments of a single $g$–$f$ erasure qubit, where $Q_{1}$ serves as the data qubit and $Q_{a}$ as the ancilla qubit for erasure detection. 

Several potential gate strategies can be used for erasure detection, including fast flux based control~\cite{Roy2023,Luo2023}, cross resonance drives~\cite{Blok2021,nguyen2024empowering,goss2022high,Fischer2023}, and four-wave-mixing (FWM) parametric interactions~\cite{Pechal2014,shirai2025high}. 
%We find the best experimental success with microwave-activated FWM parametric interaction, which will be the focus of this report.
For example, one natural approach to consider is an iSWAP gate between $|eg\rangle$ and $|0_\text{L} e\rangle$ using fast base-band flux to tune them in and out of resonance. A similar approach is also possible with flux-controlled iSWAP between $|ee\rangle$ and $|0_\text{L} f\rangle$. Furthermore, the hybridization between $|ee\rangle$ and $|0_\text{L} f\rangle$ near their avoided-crossing point leads to an ancilla frequency shift conditioned on data qubit in $\ket{e}$, allowing for microwave frequency-selective excitation of the ancilla. However, due to the weak anharmonicity of the transmon, there are often other transitions not sufficiently detuned that impacts the logical states. We find best experimental success parking the flux bias slightly away from the $|ee\rangle$-$|0_\text{L} f\rangle$ avoided crossing (the star point in Fig.~\ref{fig2}a), and driving the $|eg\rangle\leftrightarrow|0_\text{L}f\rangle$ transition to swap the erasure syndrome to the ancilla (Fig.~\ref{fig2}b). This transition is primarily FWM-like, but the final state $|0_\text{L}f\rangle$ contains some admixture of $\ket{ee}$ which enhances the gate speed. 

%One natural approach is to apply a microwave pulse to the ancilla qubit that drives the $\ket{eg} \leftrightarrow \ket{ee}\pm\ket{0_\text{L}f}$ transition, while the ancilla flux is biased at the avoided crossing formed by the $|ee\rangle$ and $|0_\text{L} f\rangle$ levels, as shown in Fig.~\ref{fig2}a. However, these transition frequency for erasure gate is close to those of the $\ket{1_\text{L}g} \leftrightarrow \ket{1_\text{L}e}$ and $\ket{0_\text{L}g} \leftrightarrow \ket{0_\text{L}e}$ transitions, which can degrade the erasure detection efficiency. 

%By exploiting the broad operating frequency range of the two-qutrit gate enabled by FWM parametric interactions, an erasure SWAP gate operation between $|eg\rangle$ and $|0_\text{L}f\rangle$ (Fig.~\ref{fig2}b) can be realized, by slightly shifting ancilla flux away from the avoided crossing point (the star point in Fig.~\ref{fig2}a) to suppress of undesired transitions. 

We calibrate this erasure SWAP operation by performing a spectroscopy of the ancilla qubit, sweeping the ancilla drive frequency $\omega_d$ for different initial states of the data qubit, of $\ket{0_\text{L}}$, $\ket{1_\text{L}}$, and $\ket{e}$, as shown in Fig.~\ref{fig2}c.  While data-qubit-dependent ancilla $\ket{g}\leftrightarrow\ket{e}$ frequency shift is apparent, driving the far-away FWM transition avoids undesirable dynamics of the logical states and provides the freedom to operate over a broad range of ancilla flux bias. An addition benefit of this erasure SWAP gate is that it re-initializes the data qubit back to the code space and therefore automatically completes an essential step for future erasure qubit operations in QEC cycles. 
%After spectroscopy of erasure SWAP gate, we performed spectroscopy of the ancillary resonator for different data-qubit states (see Fig.~\ref{fig2}d). The resonance frequency of the resonator was set to $\omega^{c}_{\mathrm{res}}/2\pi = 7.4965~\mathrm{GHz}$, together with the $\ket{f}$ state of the ancilla qubit. This configuration minimizes photon-induced dephasing of the data qubit.\bj{yingying help rewrite it}
% \yy{

Following the erasure SWAP gate implemented with a 200 ns Gaussian pulse, we perform ancilla resonator spectroscopy for different initial data-qubit states (see Fig.~\ref{fig2}d). The ancilla readout response is nearly identical for data qubit $\ket{0_\text{L}}$ and $\ket{1_\text{L}}$ states, suggesting that the ancilla readout will not dephase the logical qubit. The readout drive is set close to the ancilla resonator frequency conditioned on ancilla qubit being in $\ket{f}$ state, as indicated by the dashed line in Fig.~\ref{fig2}d. As any residual readout photons shift the ancilla qubit frequency and thus affect repeated erasure SWAP gates, this choice of readout drive frequency ensures that a large number of photons are injected only after an erasure error, while minimizing the impact on the subsequent erasure check when the data qubit remains in code space.

To characterize the performance of erasure detection which consists of the erasure SWAP gate followed by a $1.4\,\mu$s ancilla readout, we present
demodulated in-phase and quadrature (I-Q) signal from the ancilla readout for different data-qubit states. As shown in Fig.~\ref{fig2}e, the separation in the I–Q plane and the projected histograms allow us to reliably distinguish leakage from computational states. There is a 2.4\% (2.7\%) incorrect flag rate when the data-qubit state is initialized in $|0_\text{L}\rangle$ ($|1_\text{L}\rangle$). After accounting for a 0.7\% initialization error (due to thermal population, measured independently) of the data qubit, the false positive rate for $|1_\text{L}\rangle$ state is approximately 2.0\%, limited by ancilla initialization error and ancilla $T_{1}$ decay during its readout. False negative rate, where an erasure occurs but the detection fails, is approximately $6.0\%-8.0\%$, when the data qubit is initialized in $|e\rangle$. This false negative rate is mainly limited by two factors: imperfect erasure SWAP gates due to the short ancilla coherence time ($0.7$--$1\,\mu\text{s}$) away from the flux sweet spot and under FWM driving, as well as relaxation of the ancilla qubit during readout (see Supplemental Material).

%In addition to its primary function of erasure detection, our erasure SWAP gate also re-initializes the data state to the $\lvert 0_\text{L} \rangle$ state. This ``automatic reset", if further implements with resetting the ancilla qubit to ground state, allows the experiment to proceed without delay, making the protocol both resource-and time-efficient for scalable surface-code implementations. %(not detect error and reset data ground state)

\begin{figure*}[tbp]
\centering
\includegraphics[width=183mm]{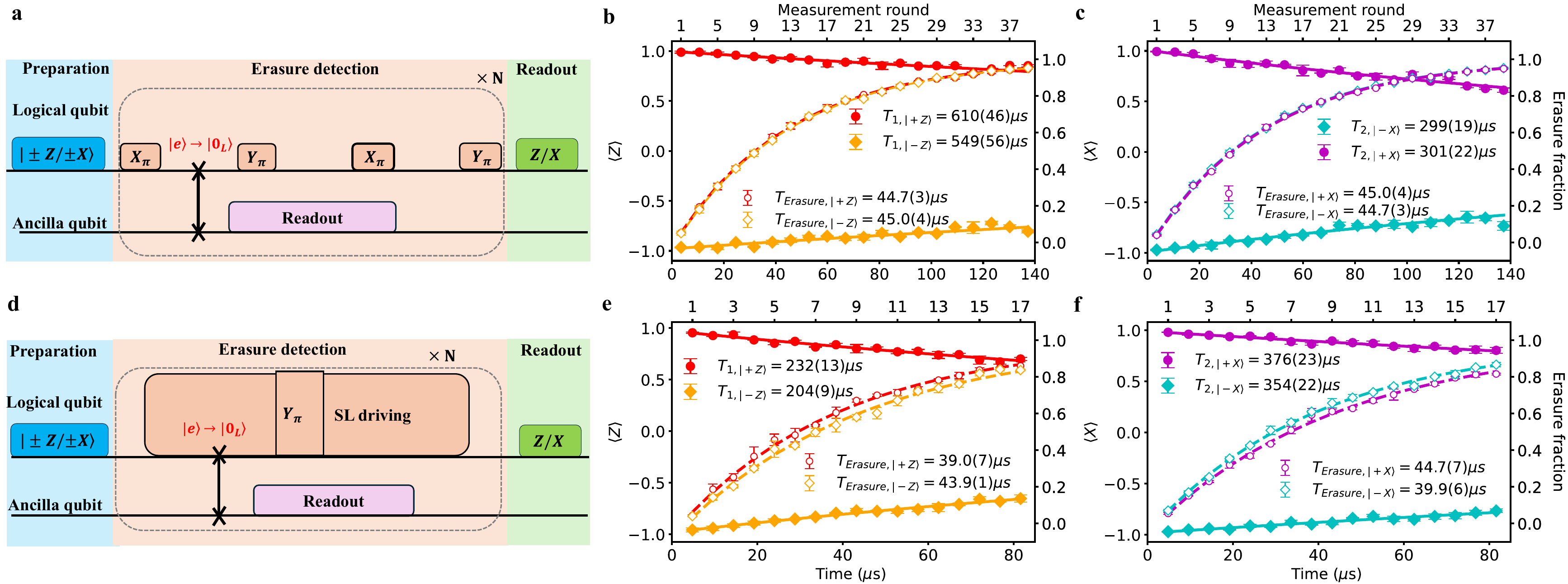}
\caption{\textbf{Measurements of coherence times of the $g$-$f$ qubit with mid-circuit erasure detection.} \textbf{a}, Diagram of XY4-based memory lifetime measurement. Each measurement round (3.52\,\(\mu\)s) incorporates the XY4 dynamical decoupling sequence and an erasure detection step. (\textbf{b,\,c}), Logical bit-flip and phase-flip times are extracted by post-selecting on the outcomes of mid-circuit erasure detection and and end-of-line measurement. Bit-flip times are measured by preparing the logical states in the Z basis,
\(|\pm Z\rangle = |0_\text{L}/1_\text{L}\rangle\), while phase-flip coherence times are measured by preparing the logical states in the X basis, \(|\pm X\rangle = (|0_\text{L}\rangle \pm |1_\text{L}\rangle)/\sqrt{2}\). Results are plotted as a function of total evolution time and measurement round. \textbf{d}, Modified protocol employing spin-locking pulses to further suppress dephasing noise, with measurements of logical Z (\textbf{e}) and X lifetimes (\textbf{f}). Extracted memory lifetimes and erasure times are obtained from exponential fits and error bars denote one standard deviation.}
\label{fig3}
\end{figure*}
\section{Memory time within logical subspace}

Building on the erasure detection scheme, we perform repeated mid-circuit erasure detection on the $g$–$f$ qubit, which monitors erasure events while largely preserving the logical states ($|g\rangle, |f\rangle$). We analyze logical lifetimes of the erasure qubit using post selection, where only trajectories with no detected leakage are retained, i.e.~any runs with an intermediate erasure are discarded. As shown in Fig.~\ref{fig3}a, each measurement round is interleaved with an XY4 dynamical decoupling sequence~\cite{maudsley1986modified}, composed of alternating $X_\pi$ and $Y_\pi$ pulses, to suppress low-frequency noise and reduce the pulse errors. These $X_\pi$ and $Y_\pi$ gates of $g$–$f$ qubit are implemented by driving the $\ket{0_\text{L}} \leftrightarrow \ket{1_\text{L}}$ two-photon transition with $80$ ns Gaussian-enveloped microwave pulses, shaped using a first-order derivative-removal-by-adiabatic-gate (DRAG) scheme~\cite{Motzoi2009}.

Frequent erasure detection is essential for the $g$–$f$ qubit due to the cascaded decay pathway, $|1_\text{L}\rangle \rightarrow |e\rangle \rightarrow |0_\text{L}\rangle$, which is characterized by the decay times $T_1^{ge}$ and $T_1^{ef}$ (with $T_1^{ge}\approx 2T_1^{ef}$). If erasure checks are too sparse, leakage error can ``seep" back into the computational subspace as an undetected logical bit-flip error. 
Each erasure detection round is performed over 3.52~$\mu$s, a duration chosen to balance the need to minimize seepage while allowing the ancilla readout photon to decay ($\kappa_r=0.3$ MHz), minimizing its effect on the subsequent erasure SWAP gate. This high cadence of mid-circuit measurements is in contrast to dual-rail qubits, whose characterization may use sparse or only end-of-the-line erasure detection~\cite{teoh2023dual,chou2024superconducting, mehta2025bias,levine2024demonstrating,huang2025logical,wills2025error} since they tend to have orders-of-magnitude lower seepage rate than erasure rate. 
%By performing erasure detection at a rate comparable to or faster than these decay processes, transient leakage events are captured and removed via post-selection, preserving logical memory lifetime. This measurement-rich approach highlights the value of high-frequency, mid-circuit monitoring for $g$–$f$ erasure qubits, in contrast to dual-rail architectures.

After discarding trajectories in which an erasure is detected via mid-circuit detection or end-of-line measurements, we extract the bit-flip lifetime within the code subspace as $549(56)\,\mu\text{s}$ and $610(46)\,\mu\text{s}$ (see~Fig.~\ref{fig3}b), for the prepared states $|+Z\rangle$ and $|-Z\rangle$, respectively, which is an order of magnitude longer than that of the bare $g$-$e$ qubit, $T_1^{ge}\sim 55\,\mu\text{s}$. The bit-flip lifetime is also much longer than the erasure lifetime ($45\,\mu\mathrm{s}$), which characterizes the timescale for detectable leakage out of the code subspace into $|e\rangle$. These timescales reflect the rate of two distinct error channels that are both important for QEC implementations. %When such an erasure is followed by re-initialization, it results a 50\% chance of a parity-flip syndrome with a known location. By contrast, the bit-flip lifetime measures Pauli errors within the code space, which produce parity-flip syndromes without a location flag and are therefore more harmful.
The erasure lifetime ($45\,\mu\mathrm{s}$) is close to the upper limit imposed by the physical decay rate $2\times T_1^{\mathrm{ef}}$ ($\sim 26\,\mu\mathrm{s}$). This should allow erasure error rates to stay comparable to the Pauli error rates of regular transmon qubits and ideally well below the fault-tolerant threshold. The ratio between erasure error and average bit flip error rate is around $(45\,\mu\text{s})^{-1}/(580\,\mu\text{s})^{-1}\approx 13$, showing that indeed the hierarchy of Pauli error rates being substantially lower than erasure error rates can be satisfied.  %dominant error affecting the $g$–$f$ qubit is the erasure error, whereas the Pauli error rates are substantially lower. 
%. While erasure errors are less damaging, both types of errors still need to be reduced to improve fault-tolerant performance. 

The bit-flip error is mainly determined by three contributions. First, double decay error ($1.0 \times 10^{-3}$) arises from the cascaded decay pathway $|1_L\rangle \rightarrow |e\rangle \rightarrow |0_L\rangle$, producing undetected logical errors. Second, false-negative–induced error ($0.9 \times 10^{-3}$) occurs when erasure events are missed, allowing leakage to propagate as logical errors. Third, pulse errors ($0.9 \times 10^{-3}$) stem from imperfections in the dynamical decoupling pulses and will be discussed in detail below. Altogether, these contributions give an expected total error of approximately $2.8 \times 10^{-3}$ per cycle, in close agreement with the measured value of ~$3.0 \times 10^{-3}$ (see Supplemental Material).

Figure~\ref{fig3}c shows coherence times of $301(22)\,\mu\text{s}$ and $299(19)\,\mu\text{s}$ when the qubit is initialized in logical $|+X\rangle$ and $|-X\rangle$ states, respectively; initializing in $|\pm Y\rangle$ states produces similar coherence times (not shown here). The observed short-time linear decays
% \bj{exponential} 
% and balanced decays 
of all six cardinal logical states towards a maximally mixed state and the symmetry between $|\pm X\rangle$ and $|\pm Y\rangle$ states suggest a simple (Makovian) Pauli logical error model consisting of a $T_{1}$ and a pure-dephasing $T_{\phi}$ process~\cite{krantz2019quantum}. To quantify the performance of our erasure-detection scheme, we compare the logical $T_{\phi}$ with the physical dephasing rate $T_\phi^{gf}$. As the erasure-qubit encoding cannot protect against physical dephasing noise, an ideal erasure qubit's lifetime would be limited by such pure-dephasing process.
% , with similar results observed for the $|\pm Y\rangle$ states. 
After subtracting the logical bit-flip's contribution, the extracted average logical pure dephasing time is $T_{\phi} = 410(30)\,\mu\text{s}$. 
% under the assumption of Markovian noise. 
For comparison, in the absence of erasure detection, the pure dephasing time between $\ket{0_L}$ and $\ket{1_L}$ is $T_{\phi}^{gf} = 440(90)\,\mu\text{s}$, which is deduced from the coherence time $T_{2,\text{XY4}}^{gf} \approx 47\,\mu\text{s}$ under similar cycles of XY4 dynamical decoupling (see Supplemental Material). These results indicate that erasure detection introduces negligible dephasing effect to the logical code space. %only a negligible reduction, within the coherence times' error bar, in logical qubit performance. 
% with the coherence time remaining close to the baseline value. 

To further suppress higher-frequency dephasing noise in $g$–$f$ qubits, we apply spin-locking~\cite{yan2013rotating,cai2012robust} sequence, a form of continuous dynamical decoupling, to the data qubit (see Fig.~\ref{fig3}d). With $4.8 \, \mu \text{s}$ erasure-detection cycles, the coherence time after post-selection is $>350\,\mu\text{s}$, with a characteristic erasure lifetime $\sim40\,\mu\text{s}$ (see Fig.~\ref{fig3}e). This improvement over the XY4 sequence demonstrates the superior noise suppression achieved with spin-locking pulse. However, the bit-flip lifetime is around $200\,\mu\text{s}$ with prepared $|\pm Z\rangle$ state, as shown Fig.~\ref{fig3}f. 
This indicates a higher rate of bit-flip errors compared to XY4, caused by the ac Stark shift of the spin-locking frequency induced by erasure detection, to which the $Z$-basis states are first-order sensitive~\cite{yan2013rotating}.

\begin{figure}[tbp]
\centering
\includegraphics[width=90mm]{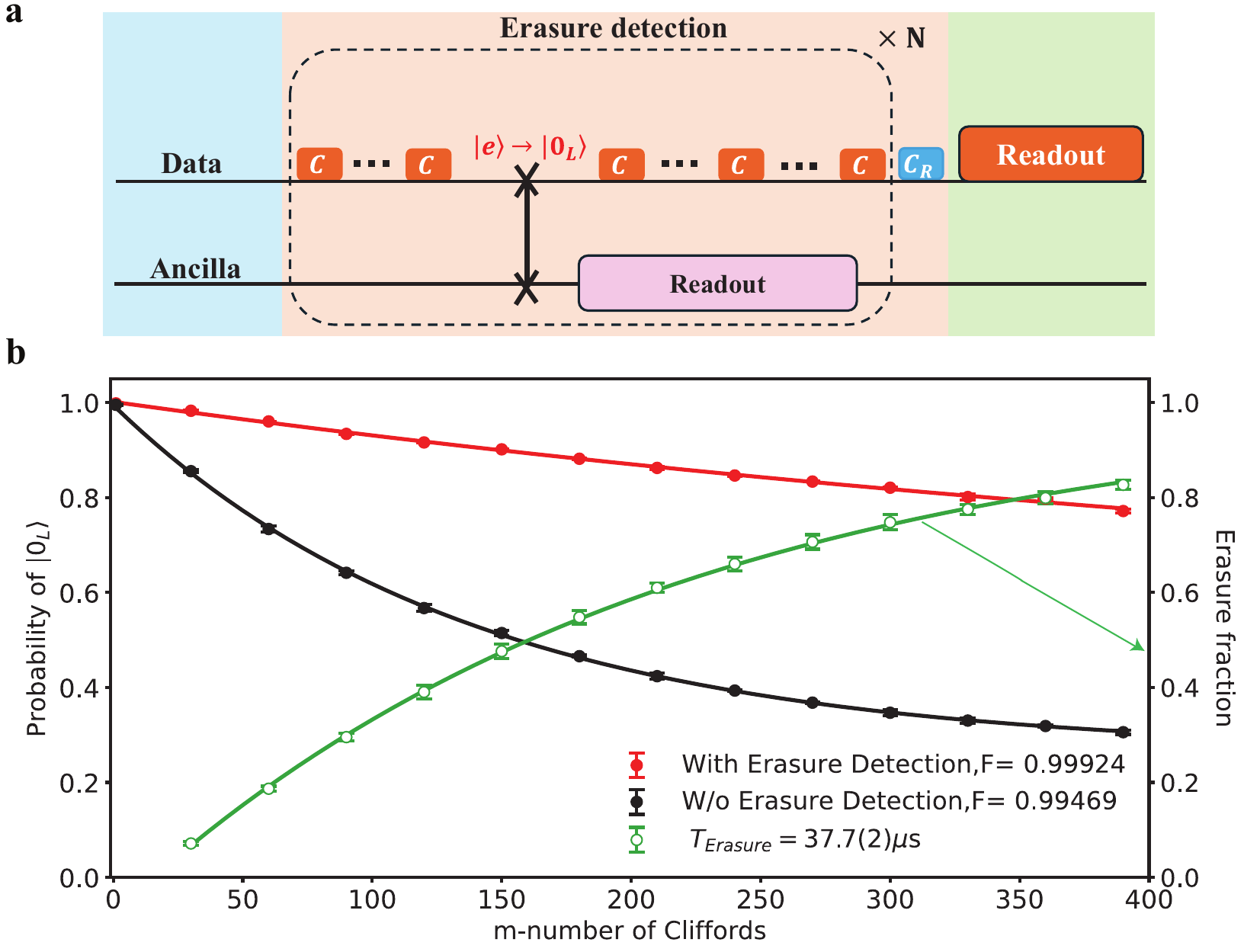}
\caption{\textbf{Single-qubit randomized benchmarking.}
\textbf{a}, Control pulse diagram of randomized benchmarking. 
\textbf{b}, Benchmarking results for single-qubit gates. The Clifford gates are compiled from a basis set of $\pi$ and $\pi/2$ rotations around the X and Y axes. Each Clifford gate consists of an average of 1.875 physical gates. A fit to the probability of $|0_\text{L}\rangle$ within the code space yields an average error per Clifford of $7.6 \times 10^{-4}$, with an erasure lifetime of $37.7(2)\,\mu\text{s}$. For the comparison, for the $g$-$f$ qubit without erasure detection, the average error per Clifford of $5.3\times 10^{-3}$ approximately seven times higher than with erasure detection.}
\label{fig4}
\end{figure}

To characterize logical single-qubit gates, we perform randomized benchmarking (RB) with mid-circuit erasure detection (see Fig.~\ref{fig4}a). The qubit undergoes a sequence of random Clifford gates followed by a final recovery Clifford. Each Clifford is decomposed into $\pi$ and $\pi/2$ rotations about the X and Y axes, with an average of 1.875 pulses per Clifford. Mid-circuit erasure detection is performed every 29 Clifford gates, with an average cycle time of $5.04\,\mu\mathrm{s}$ between successive checks. Note that no dynamical decoupling pulse is applied, as RB is robust to low-frequency dephasing noise~\cite{Epstein2014}. With post‑selection on no-erasure-error events, we obtain an average Clifford gate error of $7.6 \times 10^{-4}$ (physical gate error $4.1 \times 10^{-4}$), 
representing a seven times improvement over the raw Clifford gate infidelity of $5.3 \times 10^{-3}$ (physical gate error $2.8 \times 10^{-3}$), as shown in Fig.~\ref{fig4}b. We can infer that the gate error is dominated by leakage to $\lvert e\rangle$, thus can be effectively suppressed by post-selection via mid-circuit erasure detection. Analysis of the full error budget of Clifford gate via numerical simulations agrees well with the experimental result. (See Supplemental Material. This measurement is carried out with sub-optimal calibration of erasure detection, which in principle should perform better.) 

\begin{figure*}[tbp]
\centering
\includegraphics[width=183mm]{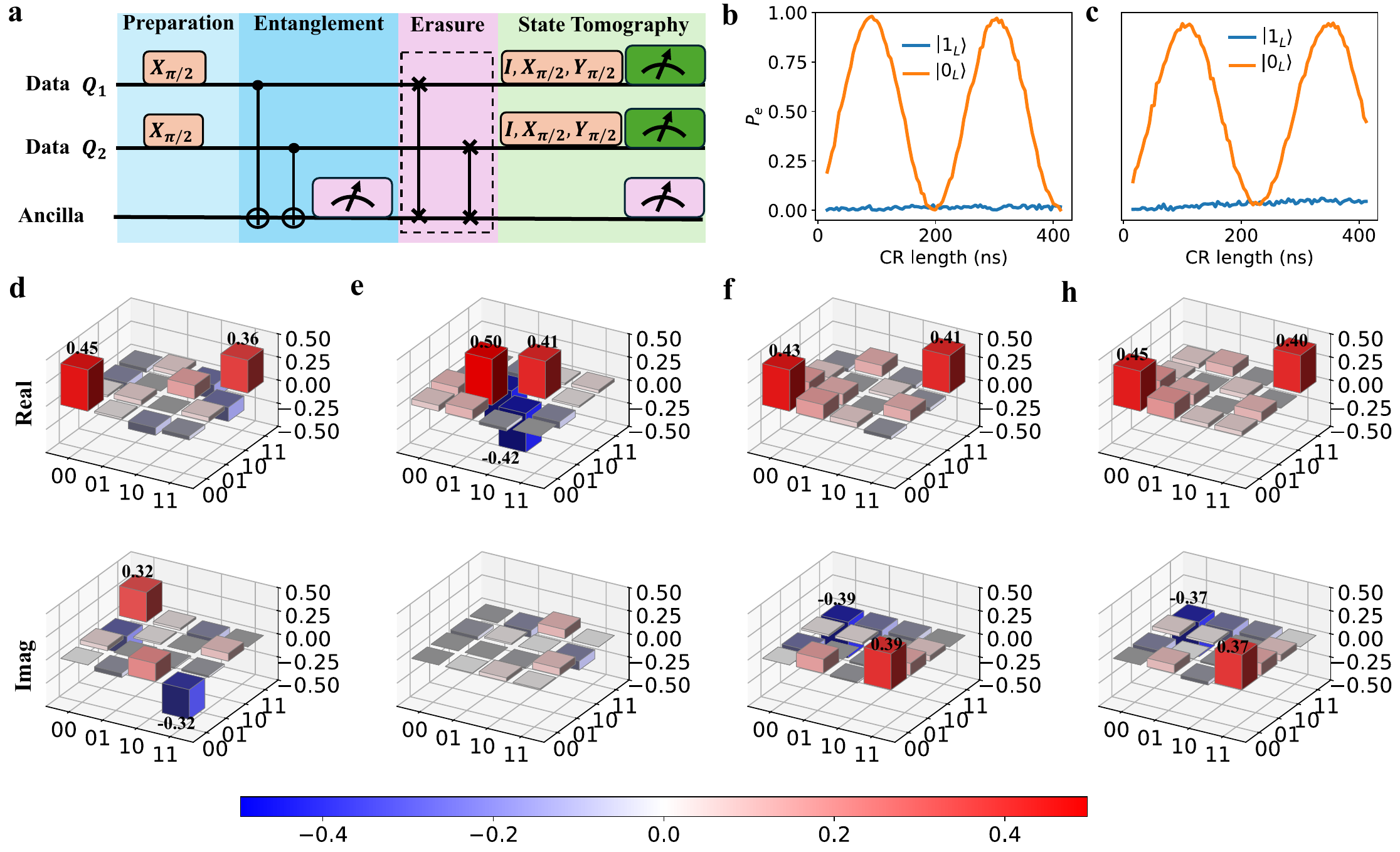}
\caption{\textbf{Parity Measurements of two $g$-$f$ erasure qubits.} \textbf{a}, Schematic of a heralded two–data-qubit entanglement experiment. After state preparation, cross-resonance CNOT gates generate a logical Bell state heralded by ancilla readout, followed by a $1.3\,\mu\text{s}$ wait and optional SWAP gate for erasure detection before two-qubit state tomography. Cross-resonance oscillations for ancilla qubit with different initial states of data qubits \( Q_1 \)  (\textbf{b}), and \( Q_2 \)  (\textbf{c}), are plotted as a function of CR pulse length. The strong dependence of the oscillation on the state of data qubit demonstrates the conditional nature of the CR interaction. Different entangled states between data qubits \( Q_1 \) and \( Q_2 \) are prepared by selecting the final state of the ancilla qubit. (\textbf{d},\,\textbf{e}), The entangled state \((|0_\text{L}0_\text{L}\rangle - i|1_\text{L}1_\text{L}\rangle)/\sqrt{2}\) and 
\((|1_\text{L}0_\text{L}\rangle - |0_\text{L}1_\text{L}\rangle)/\sqrt{2}\), are obtained by post-selecting the ancilla qubit in the \(|g\rangle\) and \(|e\rangle\) with state tomography fidelities $F=0.85(2)$ and $F=0.93(3)$, respectively. 
(\textbf{f},\,\textbf{h}), The entangled state $(|0_\text{L}0_\text{L}\rangle +i|1_\text{L}1_\text{L}\rangle)/\sqrt{2}$, is obtained by post-selecting the ancilla in $|g\rangle$, without ($F=0.89(2)$) and with ($F=0.90(3)$) erasure detection, respectively. All parity experiment data is taken during a period of slightly lower coherence performance (see Supplemental Material).
}
\label{fig5}
\end{figure*}
\section{Two-data qubit parity measurements}

%The same ancilla used for erasure detection can also implement parity measurements between data qubits, enabling efficient hardware reuse in QEC. A single ancilla mediates the interaction required to measure the joint parity of two data qubits, while remaining fully compatible with erasure-checking operations. This strategy provides a flexible, hardware-efficient pathway toward scalable quantum error-correction architectures. Here is a demonstration with two-data qubit parity measurements (see Fig.~\ref{fig5}a).

Even though the erasure detection of the $g$–$f$ qubit requires the hardware overhead of having a neighboring ancilla qubit, we propose that such ancillary resources can be fulfilled by the stabilizer-checking qubits already present in the upper-level QEC code, such as in a surface-code array. %An important feature of this architecture is that the same ancilla can be employed to both erasure detection and parity measurements, enabling a hardware-efficient approach to quantum error correction. 
This recycling of ancilla qubits for both erasure and stabilizer checking
can help reduce overall physical qubit count and control complexity, making it a practical option for scalable implementations. To demonstrate this dual functionality of ancilla qubits, we show heralded entanglement between two $g$–$f$ qubits \(Q_1\) and \(Q_2\) via a parity measurement using the same ancilla qubit \(Q_a\) previously used for erasure checking (see Fig.~\ref{fig5}a). Specifically, 
the equator states of the two data qubits are prepared by $X_{\pi/2}$ gates, followed by a pair of cross-resonance (CR) CNOT gates~\cite{Sheldon2016,Blok2021} that flip the ancilla conditioned on the data qubits in $\ket{1_L}$. The conditional Rabi oscillations of the ancilla qubit for different initial states of $Q_1$ and $Q_2$, as a function of cross-resonance drive duration are shown in Fig.~\ref{fig5}b and Fig.~\ref{fig5}c (see Supplementary Information). After a subsequent measurement of the ancilla, different Bell states of the two data qubits are heralded by post-selection on the ancilla state.
Post-selection on the ancilla states $\lvert g\rangle$ and $\lvert e\rangle$ will yield the states
$(\lvert 0_\text{L}0_\text{L}\rangle - i\lvert 1_\text{L}1_\text{L}\rangle)/\sqrt{2}$
and
$(\lvert 1_\text{L}0_\text{L}\rangle - \lvert 0_\text{L}1_\text{L}\rangle)/\sqrt{2}$.
To confirm this, we use quantum state tomography
to reconstruct the density matrix of these logical entangled states (see Figs.~\ref{fig5}d--e). The corresponding state fidelity of the generated logical
Bell state is $F = 0.85(2)$ and $F = 0.93(3)$, respectively.

To demonstrate dual-use of the ancilla qubit in the same quantum circuit, we further incorporate erasure detection on both data qubits in the heralded entanglement experiment. The reconstructed density matrices of the $\lvert g\rangle$-projected entangled state 
$(\lvert 0_\text{L}0_\text{L}\rangle + i\lvert 1_\text{L}1_\text{L}\rangle)/\sqrt{2}$ in Figs.~\ref{fig5}f--h yield fidelities of $F = 0.89(2)$ without erasure detection and $F = 0.90(3)$ when erasure detection is applied to both data qubits, indicating that erasure monitoring preserves the logical entanglement fidelity.   %There are two reasons for the absence of significant improvement. First, the erasure detection fidelity of the data qubit $Q_2$ is limited by its coherence time. Second, to make sure erasure SWAP gate is not experiencing a large frequency shift from the previous parity-checking readout, we introduced an extra $\sim 1.3\,\mu\text{s}$ waiting time before erasure gate. This waiting time is not optimal for erasure checking, since the ancilla readout photon is not fully decayed out given the small $\kappa/2\pi\sim0.3$ MHz, while the readout photon is contributing to extra dephasing to the data qubit Bell state.
The absence of significant improvement in this proof-of-principle demonstration can be attributed to the relatively short coherence time of $Q_2$ and the slow readout decay rate of the ancilla. The latter necessitates a significant delay time between the parity readout and erasure SWAP gates, during which the detriment of data-qubit dephasing errors are comparable to the benefit of erasure detection in suppressing relaxation errors. 
%\yy{There are two factors limiting the observed improvement. First, the erasure detection fidelity of $Q_2$ is limited by its coherence time. Second, a short $1.3\,\mu\text{s}$ wait is required before the erasure SWAP gate to avoid large frequency shifts from the prior parity-check readout. This wait is not long enough to fully optimize the erasure SWAP gate, yet even this short delay adds dephasing to the data qubit Bell state.}

% Those limitations can be overcome through the use of Purcell filters to get larger-linewidth ancilla resonator and optimization of circuit parameters. 
% Post-selection is restricted to the ancilla $\lvert g\rangle$ outcome because the erasure SWAP gate relies on ancilla qubit being in $\lvert g\rangle$ to work, post-selection to the ancilla $\lvert e\rangle$ outcome would also be possible with future implementation of ancilla qubit reset.
% }

% \cw{These are quite some numbers to digest, some discussions to set the expectations and compare these numbers are needed.}  %\bj{Had been moved.}   \cw{Consider this to be discussed in the outlook instead?}

\section{Discussion and Outlook} 
In summary, we experimentally demonstrated the operation of superconducting erasure qubits represented by the $\ket{g}-\ket{f}$ states of individual transmon qutrits. % superconducting qutrits using microwave-only erasure detection, avoiding fast flux control and suppressing additional logical errors. 
The mid-circuit erasure detection extends logical $Z$ and $X$ lifetimes to $500\,\mu\mathrm{s}$ and $300\,\mu\mathrm{s}$, well beyond the physical coherence times of the transmon. The logical performance is primarily limited by suboptimal ancilla readout fidelity and speed, which can be significantly improved via the use of Purcell filters~\cite{reed2010fast,Jeffrey2014,Sunada2022} and circuit parameters that mitigate measurement-induced state transitions~\cite{khezri2023measurement,dumas2024measurement,dai2026characterization,connolly2025full}. Residual dephasing noise after dynamic decoupling also contribute appreciably to the logical decoherence, whose origin and mitigation require further studies. Finally, the erasure-detected gate fidelity (with average single-qubit Clifford error of 0.076\%) is partially limited by admixture of the $\ket{e}$ states of the transmon due to its weak anharmonicity, calling for further improvement of control techniques. 

The true advantage of erasure qubits lie in their integration in a higher-level fault-tolerant QEC code such as the surface code. The original proposals of erasure-based QEC treat erasure detection as behind-the-scene operations inherent to each qubit (data qubit or checking qubit) in the QEC code with negligible spatial-temporal costs.  However, current implementations of erasure qubits such as the dual-rail qubits require not only two physical qubits to store one bit of information but also an additional ancilla to detect the erasure syndrome, resulting in a hardware-count overhead of a factor of three.  In addition to the reduction from two storage modes to one per erasure qubit, our study further suggests a potential route for additional hardware efficiency by co-designing single-qubit erasure detection with multi-qubit stabilizer checking in the upper-level QEC code. As demonstrated in our experiment, the same ancilla qutrit can be used to extract both erasure syndrome and two-qubit parity information. Further optimization of the two-qutrit gate~\cite{Roy2023,Luo2023,Blok2021} and high-fidelity three-state $\{|g\rangle, |e\rangle, |f\rangle\}$ readout~\cite{Blok2021} of the ancilla should enable simultaneous erasure and stabilizer checking in one round of syndrome measurements in the surface code. Indeed, preliminary analyses suggest that surface code of erasure qubits with this kind of delayed or reduced erasure detection can retain the benefit of steeper logical error scaling and lower fault-tolerance threshold compared to the traditional surface code~\cite{Margaret2026,liu2026achieving}.

%Erasure detection can enhance parity checks by identifying loss events while preserving entanglement fidelity, but it requires improved ancilla-qubit readout to reliably distinguish erasure from parity-flip errors using a three-state ${|g\rangle, |e\rangle, |f\rangle}$ measurement. In addition, implementing flux-assisted parametric drives enabling direct $g$–$f$ transitions for logical gate through three-wave mixing while minimizing population of the $|e\rangle$ state, could further facilitate this process by efficiently mapping erasure information onto the ancilla.

The $g$–$f$ erasure qubit demonstrated in this work is based on the industry-standard flux-tunable transmon qubits without any special design of circuit parameters, making it readily compatible with prevailing intermediate-scale superconducting quantum processors. Our experiment therefore provides a practical route toward scalable fault-tolerant architectures that explicitly exploit erasure information while maintaining modest hardware overhead.
Real-world execution of QEC codes with $g$–$f$ erasure qubits will hinge on well-tailored qutrit-qutrit gates, high-fidelity qutrit readout, reset and mitigation of leakage into higher excited states. Although we opted for microwave-activated gates between fixed-coupled transmons for both erasure extraction gates and data CNOT gates in this minimal-scale demonstration, these principle operations could be pursued with a wide variety of control strategies in various coupling architectures that may offer the necessary technical ingredients for scalability. 

%Reuse of a single ancilla for both erasure detection and parity checking eliminates the need for additional auxiliary qubits and enables more hardware-efficient quantum error-correction strategies. Together with improved multi-level readout and optimized control of higher-level transitions, these capabilities provide a practical route toward scalable fault-tolerant architectures that explicitly exploit erasure information while maintaining modest hardware overhead.

$Note$ $added$–While preparing the manuscript, we became aware of a relevant experiment~\cite{Liu2026FluxoniumErasure}, which also
demonstrates the $\ket{g}$ and $\ket{f}$ states for erasure qubit, but is based on an integer fluxonium.

\section{Acknowledgments} 
We thank M.~Pavlovich and T.~B.~Smith for helpful discussions. This work was supported by the U.S. Department of Energy, Office of Science, National Quantum In-formation Science Research Centers, Co-Design Center for Quantum Advantage under contract DE-SC0012704. Devices were fabricated and provided by the Superconducting Qubits at Lincoln Laboratory (SQUILL) Foundry at MIT Lincoln Laboratory, with funding from the Laboratory for Physical Sciences (LPS) Qubit Collaboratory. We thank MIT Lincoln Laboratory for providing the Josephson traveling wave parametric amplifier for our measurements. Y.-X.W.~acknowledges support from a QuICS Hartree Postdoctoral Fellowship. 
\bibliography{REV}

\clearpage
\onecolumngrid

\setlength{\textheight}{9.5in}
% \usepackage{braket}
% \usepackage{makecell}
% \usepackage{etoolbox,
%             mathtools}
% \usepackage{tocloft}
% \DeclarePairedDelimiterX\norm[1]\lVert\rVert{\ifblank{#1}{{\cdot}}{#1}}
% \usepackage{CJKutf8}

\newcommand{\dqb}{ {\hat s} }
\newcommand{\aqb}{ {\hat c} }

\newcommand{\resop}{ {\hat a} }
\newcommand{\resfreq}{ {\omega _{\text{res}}} }

\makeatletter
\renewcommand{\theequation}{S\arabic{equation}}
\renewcommand{\thefigure}{S\arabic{figure}}
\renewcommand{\thetable}{S\arabic{table}}
\pdfpageheight\paperheight
\pdfpagewidth\paperwidth

%% Because html converters don't know tabularnewline
\providecommand{\tabularnewline}{\\}

\renewcommand{\figurename}{Fig.}
\renewcommand{\tablename}{Table}
\gdef\@ptsize{0} % 1 for 11pt doc, 2 for 12pt
\makeatother

\pagebreak
%\widetext
	\begin{center}
	\textbf{\large Supplementary Material for ``Hardware-Efficient Erasure Qubits With Superconducting Transmon"}
\end{center}

\maketitle

\tableofcontents

\vspace{.25in}
\newpage

\section{Experimental device and setup}

\begin{figure*}[!thtbp]
	\centering
	\includegraphics[width=18cm]{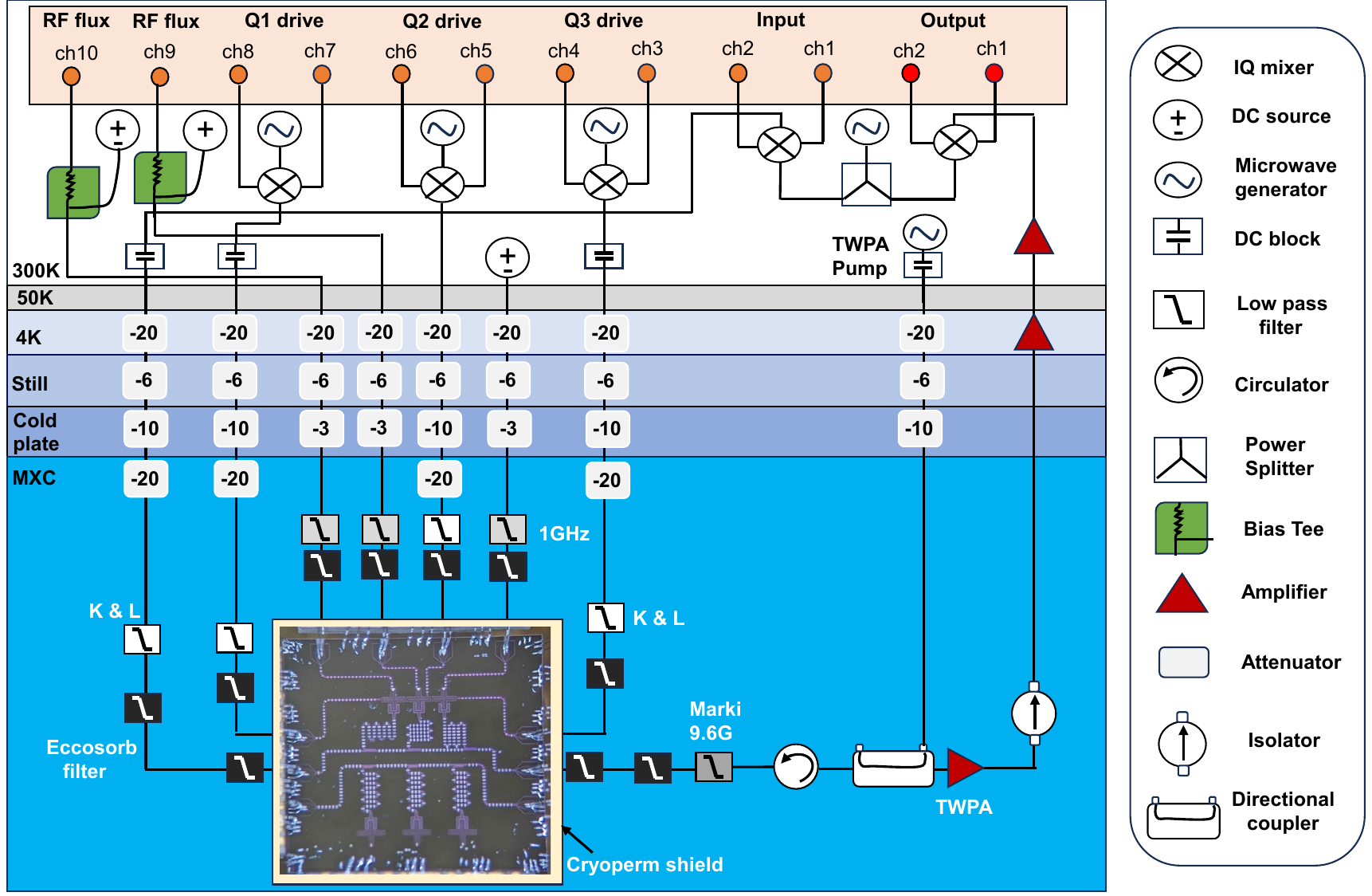}
	\caption{Cartoon depicting room-temperature microwave and cryogenic setup. }\label{Figs1}
\end{figure*}

The experimental setup incorporates a 6-qubit Xmon qubit fabricated by MIT Lincoln Laboratory, similar to that in a previous study~\cite{Chen2014}. Three qubits were used in this experiment. Room-temperature microwave and cryogenic setups are illustrated in Fig.~\ref{Figs1}. Our experiments are finished within a Bluefors LD250 dilution refrigerator, positioned at the mixing chamber (MXC) stage, maintaining a base temperature of approximately $10$ mK, with an additional layer of MXC shield to protect the qubit from IR radiation. I-Q modulation generates coherent control signals for both the readout resonators and the transmons. We utilize a Traveling Wave Parametric Amplifier (TWPA) at the MXC stage and a High-Electron-Mobility Transistor (HEMT) amplifier at the $4$ K stage. To shield the experiment from external magnetic fields, it is housed in an Amuneal can. We paint the inner surface of the Amuneal can with black painting materials (Stycast 1266, Carbon Black and Silicon Carbide)
to further suppress the effect of IR radiation. Radio-frequency (RF) lines are filtered using a K\&L $12$ GHz low-pass filter, a Marki $9.6$ GHz low-pass filter, and Eccosorb low-pass filters. A $10$ MHz reference signal is employed to phase-lock all radio-frequency instruments, including the ADC. For fast data acquisition, we use the OPX+ system in conjunction with microwave sources (Lab Brick and Windfreak) to generate microwave pulses. Magnetic flux control is achieved using a bias tee (ZFBT-4R2GW-FT+) that combines an RF flux signal with a DC bias supplied by a stabilized current source  (Yokogawa $7651$).  The control and coherent parameters of the three-qubit device in this experiment are listed in Table~\ref{table1}.

\begin{table}[t]
\caption{Qubit parameters during two seperate fridge cooldowns( A, B)}\label{table1}
\label{tab:three_qubit_transposed}
\begin{ruledtabular}
\begin{tabular}{ccccccc}
Parameter & $Q^{A}_{1}$ & $Q^{A}_{a}$ & $Q^{A}_2$& $Q^{B}_{1}$ & $Q^{B}_{a}$ & $Q^{B}_2$ \\
\hline
ge frequency (zero-flux point) $\omega/2\pi$ (GHz) & 5.239 & 6.082 & 5.252 & 5.182 & 6.020 & 5.192 \\
ge frequency (operating   point) $\omega_{\text{bias}}/2\pi$ (GHz) & 5.239 & 5.472 & 5.252 & 5.182 & 5.410 & 5.192 \\
Coupling strength $g/2\pi$ (MHz) & 12.5 & \ & 11.5 & \ & \ & \ \\
Readout frequency $\omega_r/2\pi$ (GHz) & 7.153 & 7.496 & 6.933 & 7.153 & 7.496 & 6.933\\
Readout decay $\kappa_r/2\pi$ (MHz) & 0.21 & 0.30 & 0.30 & 0.21 & 0.30 & 0.30\\
Dispersive coupling $\chi/2\pi$ (MHz) & 1.0 & 1.1 & 1.2 & 1.0 & 1.1 & 1.2 \\
Amharmonicity $\alpha_q/2\pi$ (MHz) & 180 & 183 & 180 & 180 & 183 & 180  \\
 $\ket{e} \rightarrow \ket{g}$ decay $T^{ge}_1 (\mu s)$ & 30-70 & 15-20 & 15-25 & 40-80 & 10-15 & 20-30 \\
 $g-e$ Ramsey  $T^{ge}_2 (\mu s)$ & 20-30 & 10-25 & 10-25 & 40-80 & \ & \ \\
 $g-e$ Echo  $T^{ge}_{E} (\mu s)$ & 20-40 & 17 & 20-30  & 50-90 & 15-18 & 25 \\
$\ket{f} \rightarrow \ket{e}$ decay $T^{ef}_1 (\mu s)$& 10-30 & 8 & 10-20 & 15-35 & 5-10 & 15\\
$g-f$ Ramsey  $T^{gf}_2 (\mu s)$ & 20-30 & \ & 10-15 & 20-40 & \ & \ \\
\end{tabular}
\end{ruledtabular}
\end{table}

\section{Qutrit Readout} 

\begin{figure*}[!tbh]
    \centering
    \includegraphics[width=15cm]{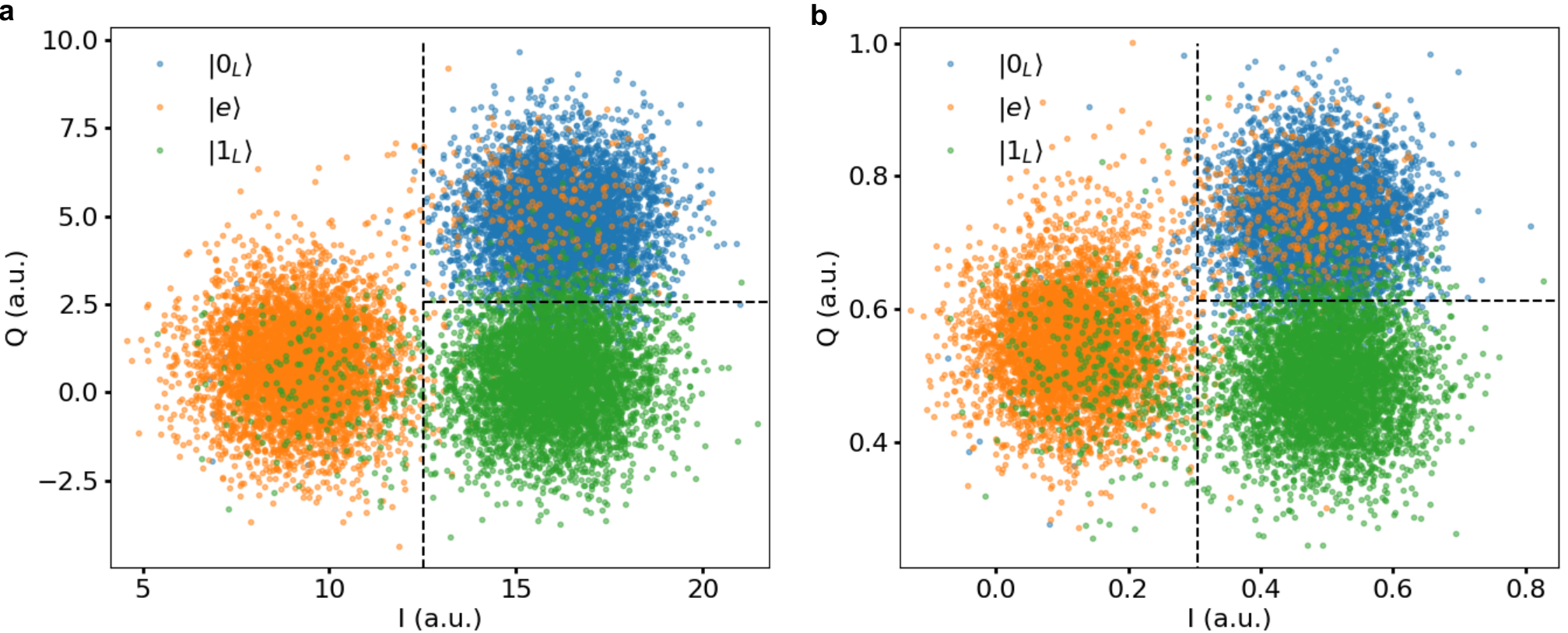}
    \caption{Distribution of the I–Q cloud obtained from readout measurements of (\textbf{a}) $Q_{1}$ and (\textbf{b}) $Q_{2}$ for different prepared states. The two dashed lines indicate the I and Q thresholds used to distinguish the $\ket{0_{\text{L}}}$, $\ket{1_{\text{L}}}$, and $\ket{e}$ states. }
    \label{fig:gef}
\end{figure*}

To accurately determine qutrit state populations, we employ a qutrit readout calibration matrix to suppress thermal excitations and ensure initialization in the ground state \( \lvert 0_{\text{L}} \rangle \). Readout errors are then mitigated using a three-state calibration matrix for the \( \lvert 0_{\text{L}} \rangle \), \( \lvert e \rangle \), and \( \lvert 1_{\text{L}} \rangle \) states. To construct this matrix, the qutrit is prepared in each of the basis states \( \lvert 0_{\text{L}} \rangle \), \( \lvert e \rangle \), and \( \lvert 1_{\text{L}} \rangle \), and repetitive single-shot measurements are performed for $Q_{1}$ and $Q_2$ as shown Fig.~\ref{fig:gef}.
In the basis states \( \lvert g \rangle \), \( \lvert e \rangle \), and \( \lvert f \rangle \), the  readout calibration matrix $M_{Q_1}$ and $M_{Q_2}$ of data qubits $Q_1$ and $Q_2$ are given by,
\begin{equation}
M_{Q_1} =
\begin{bmatrix}
0.941 & 0.017 & 0.042 \\
0.043 & 0.951 & 0.006 \\
0.050 & 0.040 & 0.91
\end{bmatrix}, \,
M_{Q_2} =
\begin{bmatrix}
0.944 & 0.021 & 0.035 \\
0.076 & 0.916 & 0.008 \\
0.048 & 0.077 & 0.874
\end{bmatrix} 
\end{equation}
where each column corresponds to the probability distribution of measured qutrit assignments conditioned on the preparation of a given qutrit states of data qubits.

\section{Erasure checking with fast flux pulses} 

\begin{table}[h!]
\centering
\begin{tabular}{ccc|cc}
\hline
&\multicolumn{2}{c|}{$Q_{1}$-$Q_a$} & \multicolumn{2}{c}{$Q_{a}$-$Q_2$} \\
Level Crossing & Freq (MHz) & $g$ (MHz) & Freq (MHz) & $g$ (MHz) \\
\hline
$|e g\rangle - |0_{\text{L}} e\rangle$ & 5181.99 & 25.00 & 5192.00 & 23.00 \\
$|1_{\text{L}} g\rangle - |e e\rangle$ & 5002.05 & 35.33 & 5011.99 & 32.50 \\
$|1_{\text{L}} g\rangle - |0_{\text{L}} f\rangle$ & 5183.48 & 3.38 & 5193.50 & 2.87 \\
$|e e\rangle - |0_{\text{L}} f\rangle$ & 5364.98 & 35.32 & 5375.00 & 32.50 \\
$|h g\rangle - |e f\rangle$ & 5003.53 & 5.79 & 5013.49 & 4.92 \\
$|h g\rangle - |0_{\text{L}} h\rangle$ & 5185.00 & 0.18 & 5195.00 & 0.14 \\
$|1_{\text{L}} e\rangle - |e f\rangle$ & 5185.00 & 49.82 & 5195.00 & 45.86 \\
$|1_{\text{L}} e\rangle - |0_{\text{L}} h\rangle$ & 5366.53 & 5.79 & 5376.50 & 4.92 \\
$|e f\rangle - |0_{\text{L}} h\rangle$ & 5548.05 & 43.23 & 5558.00 & 39.78 \\

% 10 - 01 5191.99879787229 MHz, g: 23.000058476417507 MHz
% 20 - 11 5011.991807358394 MHz, g: 32.502382458251304 MHz
% 20 - 02 5193.501981766663 MHz, g: 2.8692670155560336 MHz
% 11 - 02 5375.001214797152 MHz, g: 32.50253119493027 MHz
% 30 - 12 5013.493174201894 MHz, g: 4.919978044463278 MHz
% 30 - 03 5195.00245194358 MHz, g: 0.13800150437600678 MHz
% 21 - 12 5195.00245194358 MHz, g: 45.86232132942314 MHz
% 21 - 03 5376.499666150328 MHz, g: 4.919985830591031 MHz
% 12 - 03 5558.008269870165 MHz, g: 39.78381146389802 MHz

\hline
\hline
\end{tabular}
\caption{%Transitions level crossings between data qubit qubit and ancilla qubit, plus the ancilla frequencies and the coupling strength $g$ of the level crossings.
Data qubit ($Q_{1}$) and ancilla qubit ($Q_a$) system level crossings. Ancilla frequencies at which the bare joint states of the data–ancilla system are degenerate. 
The parameter $g$ denotes the coupling strength extracted from the resulting avoided level crossings.}\label{table2}
\end{table}

The experiment implements an erasure gate between a data qubit and an ancilla, such that the ancilla is excited if and only if the data qubit occupies the $|e\rangle$ state, as mentioned in the main text. One can also use flux-based iSWAP and controlled-Z gates for erasure detection~\cite{barends2019diabatic}. However, this approach introduces certain issues, which are discussed in the following.

\begin{figure*}[!thtbp]
	\centering
	\includegraphics[width=18cm]{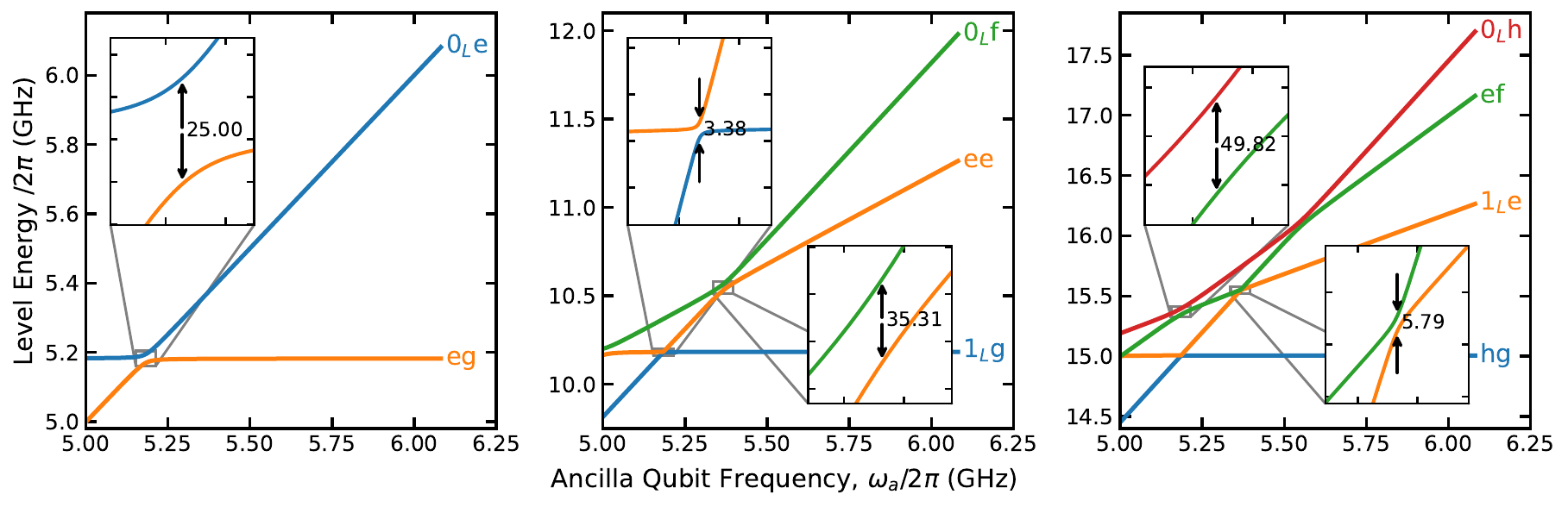}
	\caption{Energy levels diagram of two qubit ($Q_1$, $Q_a$) interacting system, as a function of ancilla qubit's frequency ($Q_{a}$. Inset figures illustrate zoomed in level crossings}\label{Figs2}
    
\end{figure*}

For erasure detection with the fast flux pulse, an iSWAP gate between two capacitively coupled transmons of data and ancilla qubits is implemented by dynamically tuning ancilla qubit with a shaped flux pulse. In the zero-flux point, the qubits are strongly detuned so that their static exchange interaction $g$ is effectively suppressed. Applying a fast flux pulse shifts the transition frequency of the ancilla transmon, $\omega_{a}(t)$, and brings it into resonance with the data transmon, thereby activating the exchange interaction. When the qubits satisfy the resonant condition, it will produce the iSWAP gate under the $\ket{eg} \leftrightarrow \ket{0_{\text{L}}e}$ transition. However, in practice, the transmon is not a strict two-level system but a weakly anharmonic multi-level system, resulting in the presence of non-computational subspaces, which are shown in the Table~\ref{table2} and Fig.~\ref{Figs2}. 
Specifically, these transition frequency $\ket{eg} \leftrightarrow \ket{0_{\text{L}}e}$ for erasure gate is close to those of the $\ket{1_{\text{L}}g}\leftrightarrow \ket{0_{\text{L}}f}$ transitions, which can reduce the erasure detection efficiency.

A controlled-$Z$ (CZ) gate between the ancilla and data qubits is implemented by dynamically tuning the ancilla frequency into an avoided-level crossing with the $\ket{ee} \leftrightarrow \ket{0_{\text{L}}f}$ transition. As the ancilla approaches this crossing, the $\ket{ee}$ state hybridizes with $\ket{0_{\text{L}}f}$, acquiring an energy shift. By selecting an appropriate interaction time, this shift produces a conditional dynamical $\pi$ phase while leaving the other computational states nearly unaffected.
However, the nearby transition $\ket{1_{\text{L}}e} \leftrightarrow \ket{0_{\text{L}}h}$ can reduce the fidelity of erasure detection due to its proximity in frequency to the target transition (see Table~\ref{table2} and Fig.~\ref{Figs2}).

% \begin{table}[h]
% \centering
% \begin{tabular}{l l }
% \hline
% \textbf{SPAM error} & \textbf{Parameterization} \\
% \hline
% Thermal population of data qubit
% & $0.73\%$
% \\[10pt]

% Initialization error of data qubit 
% & $1.03\%$
% \\[10pt]

% Data readout error 
% $|g\rangle\rightarrow|e\rangle$
% &  $0.07\%$

% \\[12pt]
% Data readout error 
% $|e\rangle\rightarrow|g\rangle$
% &  $2.2\%$
% \\[12pt]
% Thermal population of ancilla qubit
% & $0.94\%$
% \\[10pt]

% \hline
% \end{tabular}
% \end{table}

\section{Error analysis of memory lifetime}

\subsection{False positive and negative rate}

With preparing different states of data qubit, i.e., $|0_{\text{L}}\rangle$, $|1_{\text{L}}\rangle$ and $|e\rangle$ state, we obtain the probabilities of different outcomes for ancilla readout, which was shown in Fig. 2(e) of main text. From that, 
the false positive rates of $|0_{\text{L}}\rangle$ and $|1_{\text{L}}\rangle$ states are given by, $\epsilon^{|0_{\text{L}}\rangle}_{fp}=P(0_{\text{L}}e|0_{\text{L}}g)+P(ee|0_{\text{L}}g)+P(1_{\text{L}}e|0_{\text{L}}g)\approx2.4\%$ and $\epsilon^{|1_{\text{L}}\rangle}_{fp}=P(0_{\text{L}}e|1_{\text{L}}g)+P(ee|1_{\text{L}}g)+P(1_{\text{L}}e|1_{\text{L}}g)\approx2.7\%$, which includes the contribution caused by 0.7\% thermal population of data qubit. 

The false-negative rate was mainly limited by the coherence of the ancilla qubit, which was operated approximately 600 MHz away from the zero-flux point. The $g$--$e$ Ramsey coherence time, $T_2^{ge}$, and the $g$--$e$ Ramsey echo coherence time, $T_{2E}^{ge}$, were measured to be $\sim 0.42\,\mu\text{s}$ and $\sim1.1\,\mu\text{s}$, respectively, while the driven coherence under four-wave mixing driving was approximately 0.8\,$\mu\text{s}$. To reduce measurement-induced state transitions, the ancilla readout length was extended to about 1.4\,$\mu\text{s}$ to maintain sufficient signal-to-noise ratio. However, this longer readout time increases errors due to relaxation of the ancilla qubit during readout measurement. In addition, the false negative rate exhibits fluctuations due to coherence variations and instability in the erasure SWAP gate frequency induced by flux noise.

\begin{figure}[!thtbp]
	\centering
	\includegraphics[width=8cm]{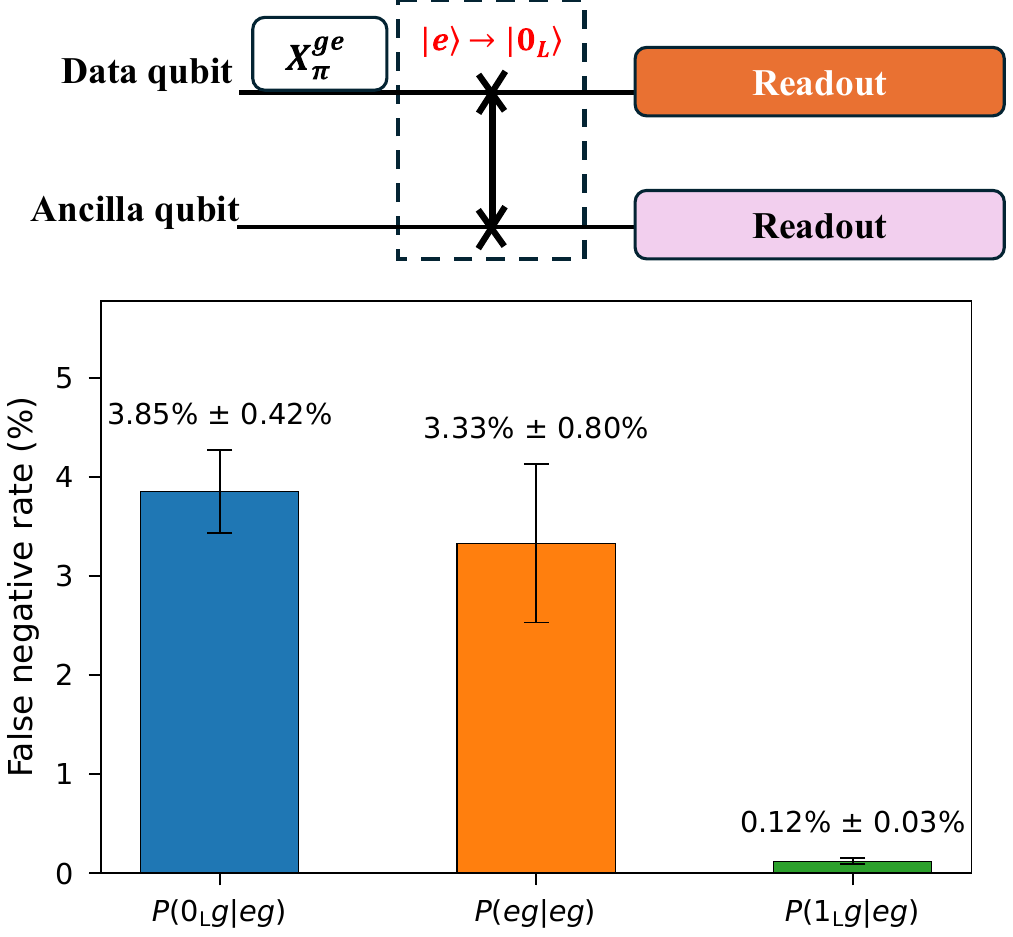}
	\caption{Experimental data for false-negative rates from different coontributing channels. Top: pulse sequence for preparing and measuring false-negative events. Bottom: False-negative rates, with data qubit initialized in $\ket{e}$ state and measured in $|0_L\rangle$, $|e\rangle$, and $|1_L\rangle$ states.}\label{Figsfn}
    
\end{figure}

To understand the false-negative effects in our XY4-based mid-circuit erasure detection protocol, we further quantify the false-negative probabilities $P(0_{\text{L}}g|eg)$, $P(1_{\text{L}}g| eg)$, and $P(eg|eg)$. With the data qubit initialized in the $\ket{e}$ state, we perform simultaneous readout of both the data and ancilla qubits following the erasure SWAP gate. The results are shown in Figs.~\ref{Figsfn}, after correcting for the data-qubit thermal population and pulse errors (approximately $1\%$).
Therefore, the total false negative rate is given by, $\epsilon_{fn}=P(0_{\text{L}}g|eg)+P(1_{\text{L}}g|eg)+P(eg|eg)\approx7.3(0.9)\%$. 

\subsection{Error budget of bit-flip lifetime}
Here, we report the specific error budget for the average bit-flip lifetime, which discussed in the main text. Decay errors predominantly induce leakage to the $\ket{e}$ state in the $g$–$f$ erasure qubit via $\ket{f} \rightarrow \ket{e}$ relaxation. The per-cycle leakage can be approximated as $\epsilon_{\text{l}} = 1 - \exp(-t_{\rm cycle} / 2T_1^{ef}) \approx 6.8\%$ for our XY4 scheme, where the duration of a measurement round is $t_{\rm cycle} = 3.52\,\mu\text{s}$. Although this leakage is, in principle, detectable using the erasure detection protocol, several mechanisms still degrade the logical qubit performance, as summarized in Table~\ref{table3}. First, cascaded decay events contribute at second order, yielding a short-time contribution of approximately $\frac{t_{\rm cycle}^2}{8 T_1^{ef} T_1^{ge}} \approx 1\times 10^{-3}$. Since such double-decay events cannot be detected by the erasure detection scheme, they directly induce logical errors and therefore set a fundamental limit on the bit-flip lifetime. Second, detection-related errors are dominated by false-negative events arising from imperfect erasure detection in the XY4 protocol. These contributions scale as $\epsilon_{\text{l}}$ multiplied by the corresponding conditional misclassification probabilities. For the $|+Z\rangle$ state, the dominant contribution is $2\epsilon_{\text{l}} P(0_{\text{L}}g|eg)/3 \approx 1.7\times 10^{-3}$ (note that around 1/3 of $P(0_{\text{L}}g|eg)$ comes from the ancilla readout misassignment which could be captured in the next erasure checking), while an additional term $\epsilon_{\text{l}} P(eg |eg)\frac{1 - \exp(-t_{\rm cycle} / T_{1}^{eg})}{2} \approx 7\times 10^{-5}$ accounts for decay occurring before the next erasure detection cycle. For the $|-Z\rangle$ state, the corresponding contribution is $\epsilon_{\text{l}} P(1_{\text{L}} g|eg) \approx 8\times 10^{-5}$. Averaging over logical states of $|\pm Z\rangle$, the resulting bit-flip error from erasure detection is $\sim 0.9 \times 10^{-3}$.
Third, pulse errors in the XY4 sequence, characterized via randomized benchmarking, contribute approximately $0.9 \times 10^{-3}$.

\begin{table}[t]
\centering
\caption{Error contributions per cycle for the $|+Z\rangle$ and $|-Z\rangle$ logical states.}\label{table3}
\begin{tabular}{lcc}
\hline
\textbf{Error source} 
& \textbf{$|+Z\rangle$} 
& \textbf{$|-Z\rangle$} \\
\hline\
Leakage ($\epsilon_{\text{l}}$) 
& $1 - \exp\!\left(-\dfrac{t_{\mathrm{cycle}}}{2T_{1}^{ef}}\right)$ 
& $1 - \exp\!\left(-\dfrac{t_{\mathrm{cycle}}}{2T_{1}^{ef}}\right)$ \\
Cascaded decay loss 
& $\sim\dfrac{t_{\mathrm{cycle}}^2}{8T_{1}^{ef}T_{1}^{ge}}$ 
& $\sim\dfrac{t_{\mathrm{cycle}}^2}{8T_{1}^{ef}T_{1}^{ge}}$ \\

False negative ($P(0_{\text{L}}g|eg)$) probability 
& $\sim2\epsilon_{\text{l}} P(0_{\text{L}}g|eg)/3$ 
& $0$ \\

False negative ($P(1_{\text{L}}g|eg)$) probability   
& $0$ 
& $\epsilon_{\text{l}} P(1_{\text{L}}g|eg)$ \\

False negative ($P(eg|eg)$) probability 
& $\epsilon_{\text{l}} P(eg|eg) (1 - e^{-t_{\mathrm{cycle}}/T_{eg}})/2$ 
& $0$ \\

Logical gate error
& $2.34\times10^{-4}$ 
& $2.34\times10^{-4}$ \\
\hline
\end{tabular}
\end{table}

\begin{figure}[!thtbp]
	\centering
	\includegraphics[width=10cm]{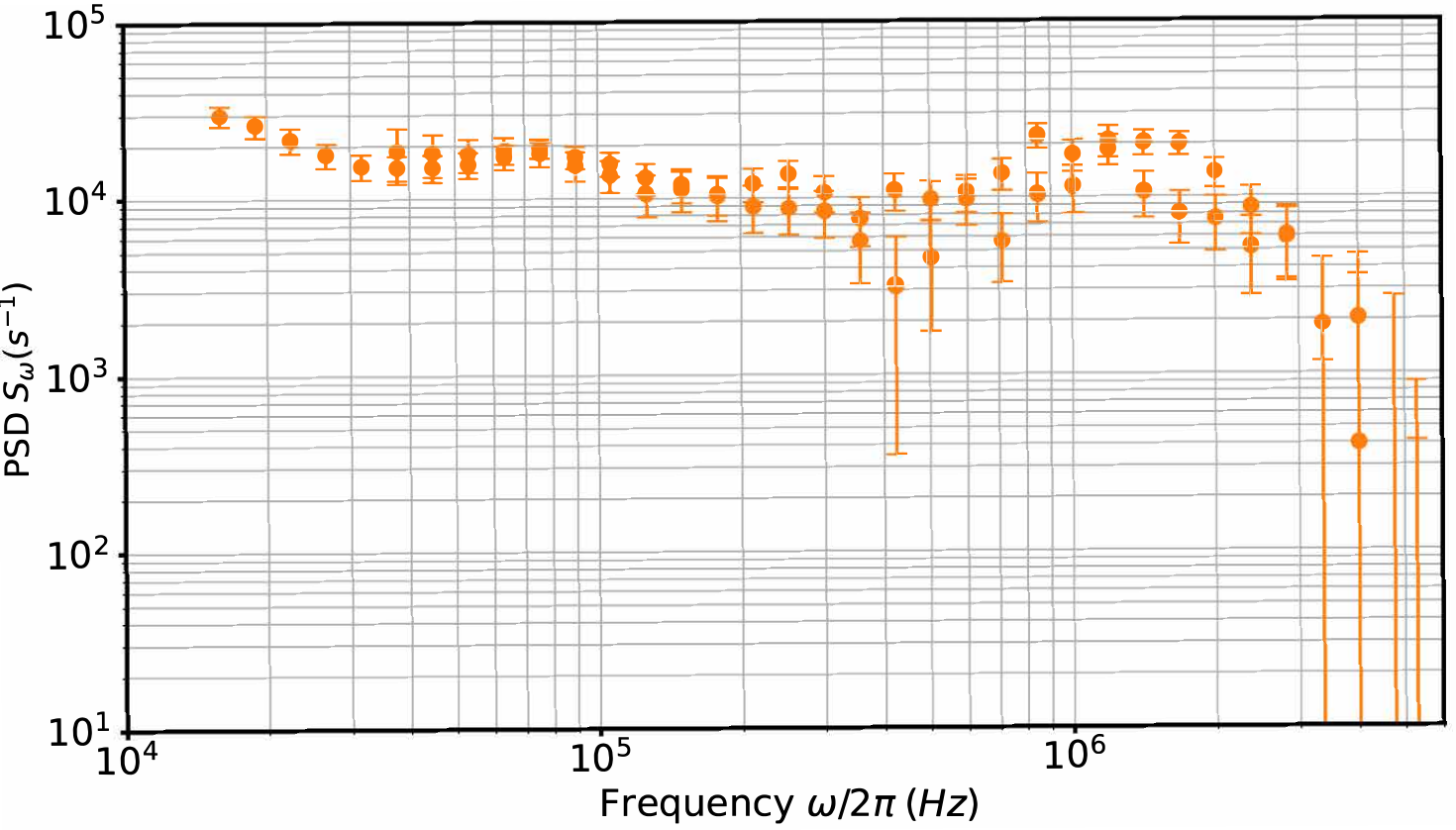}
	\caption{Noise spectral density of $g$--$f$ qubit, measured with forward and backward frequency sweeps.}\label{gfpsd}
\end{figure}

\subsection{Error budget of phase-flip lifetime}
\subsubsection{Dephasing noise in $g$--$f$ qubit}
We next discuss the dephasing noise in the $g$--$f$ erasure qubit, which, unlike dual-rail erasure qubit~\cite{Campbell2020,teoh2023dual,levine2024demonstrating}, does not benefit from a large energy gap for the coherence protection. Here we employed a spin-locking pulse sequence~\cite{yan2013rotating} to measure the noise spectral density (PSD), which characterizes coherence fluctuation. The spin-locking pulse sequence is similar to that shown in Fig.~3d in the main text. The relaxation time $T^{gf}_{1\rho}$ in the rotating frame was measured over a range of spin-locking Rabi frequencies. Under a weak resonant drive, the noise power spectral density, $S_\omega$ of the $gf$ qubit is extracted via
\begin{equation}
\frac{1}{T^{gf}_{1\rho}} = \frac{S_\omega}{2} + \frac{1}{2T^{ef}_{1}}.
\end{equation}
The extracted PSD (Fig.~\ref{gfpsd}) shows that dephasing noise is significantly suppressed at higher spin-locking Rabi frequencies. This is consistent with the improved coherence observed in the memory experiments when spin-locking pulses are applied. Alternatively, the dephasing noise can also be suppressed by implementing dynamical decoupling pulses, provided that high-fidelity $\pi$ pulses are available, as will be discussed in the following subsection.
\begin{figure}[!thtbp]
	\centering
	\includegraphics[width=10cm]{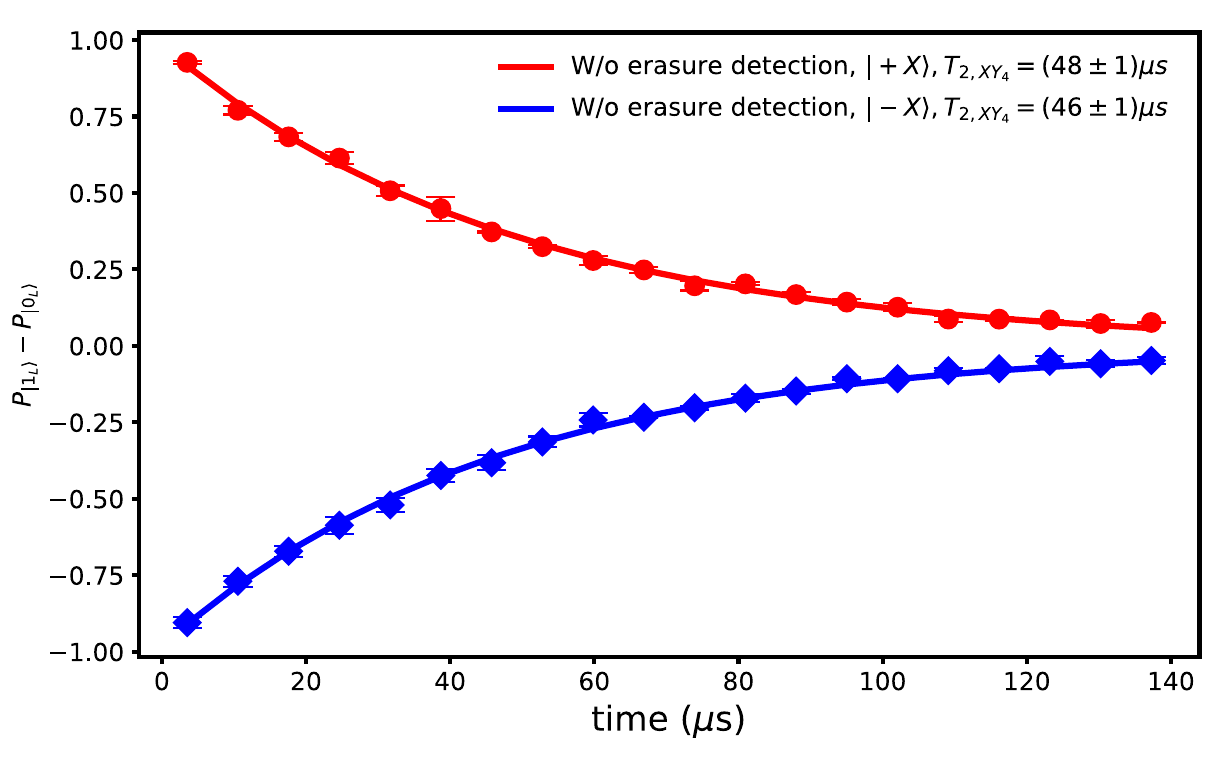}
	\caption{Coherence measurements with the XY4 pulse sequence (no erasure detection).The population difference \(P_{|1_{\rm L}\rangle} - P_{|0_{\rm L}\rangle}\) is plotted as a function of evolution time for the logical states \(|+X\rangle\) (red circles) and \(|-X\rangle\) (blue diamonds). The data points represent experimental measurements, while the solid lines show exponential fits to extract the coherence times \(T_2^{XY4}\). Extracted values are \(T_2^{XY4} = (48 \pm 1)\,\mu s\) for \(|+X\rangle\) and \(T_2^{XY4} = (46 \pm 1)\,\mu s\) for \(|-X\rangle\).}\label{XY4}
    
\end{figure}
\subsubsection{Phase-flip coherence time with erasure detection}
To better understand noise suppression under dynamical decoupling pulses, we measured XY4 coherence times without erasure detection using $|+X\rangle$ and $|-X\rangle$ initial states, as shown in Fig.~\ref{XY4}. The average coherence time is 
$T_{2,\text{XY4}} \approx 47\,\mu\text{s}$, with a relaxation time of $T^{ef}_{1}\approx26\,\mu\text{s}$. 
As discussed in the main text, the short-time linear decay toward a maximally mixed state, together with the symmetry between $|\pm X\rangle$ and $|\pm Y\rangle$, suggests a Markovian Pauli error model characterized by $T_1$ relaxation and pure dephasing $T_\phi$~\cite{krantz2019quantum}. We extract a pure-dephasing coherence time of $T_\phi \approx 440\,\mu\mathrm{s}$ in the absence of erasure detection. Incorporating the logical bit-flip time under erasure detection yields a predicted average coherence time of $T_2 \approx 320\,\mu\mathrm{s}$, slightly exceeding the measured value; we expect this discrepancy to be due to additional dephasing induced by imperfect erasure detection.

\subsubsection{Ancilla-readout-induced dephasing noise}
Here we consider the dispersive coupling between the data transmon and the ancilla resonator as mediated by the ancilla transmon, in the presence of resonator readout drive at frequency $\omega _{\mathrm{dr}}$ with drive strength $f_{\text{dr}}(t)$. We start from the general total system Hamiltonian, $\hat H _{\mathrm{tot}} 
=  \hat H _{\mathrm{0}} +\hat H _{\mathrm{int}} + \hat H _{\mathrm{dr}}  $, where  
\begin{equation}\label{s2}
\begin{aligned}
\hat H_{\mathrm{0}} 
&= \omega_{\text{s}} \dqb^\dag \dqb 
- \frac{\alpha_{\text{s}}}{2}\dqb^\dag \dqb^\dag \dqb \dqb+\omega_{\text{a}} \aqb^\dag \aqb 
- \frac{\alpha_{\text{a}}}{2} \aqb^\dag \aqb^\dag \aqb \aqb  
+ \omega_{\text{r}} \resop^\dag \resop , \\
\hat H_{\mathrm{int}}   
&= g(\dqb + \dqb^\dag)(\aqb + \aqb^\dag)
+ g_{\text{r}} (\aqb + \aqb^\dag)(\resop + \resop^\dag), \\
\hat H_{\mathrm{dr}} &= f_{\text{dr}}(t) e^{i \omega_{\mathrm{dr}} t} \resop 
+ \text{H.c.} , \\
\end{aligned}
\end{equation}
In Eq.~\eqref{s2}, $\dqb$, $\aqb$, and $\resop$ denote the annihilation operators corresponding to the data qubit, the ancilla qubit, and the ancilla readout resonator, respectively. The coupling strength between the ancilla transmon and resonator is chosen such that their dispersive shift $\chi_{a,\text{bias}} $ matches the experimentally measured value (at bias/idling point):
\begin{align}
\chi_{\text{a},\text{bias}} = - \frac{2 g_{\text{r}}^{2} \alpha_a }{ (\omega^{\text{bias}}_\text{a} - \omega_{\mathrm{r}}) (\omega_{\text{a},\text{bias}} -\alpha_\text{a} - \omega_{\mathrm{r}} ) }
\approx - 0.53 ~\text{MHz}
.
\end{align}
which yields \( g_{\text{r}}/2\pi \approx 103~\mathrm{MHz} \). The dispersive coupling between the data transmon and the ancilla resonator arises from two contributions: the  dispersive interaction between the ancilla transmon and its readout resonator, and the hybridization induced by the \( \ket{1_{\text{L}} g} \leftrightarrow \ket{ee} \) transition. An analytical approximate estimate of the effective \( g\!-\!f \) dispersive shift is 
\begin{align}
\chi_{\text{sa},\mathrm{eff}} \sim \chi_{\text{a},\mathrm{bias}} 
\left( \frac{\sqrt{2}\, g}{\omega_\text{q} - \alpha_\text{q} - \omega_{\text{a},\mathrm{bias}}} \right)^2 
\approx -1.0~\mathrm{kHz}
,
\end{align}
contributing to an effective dispersive term coupling the logical qubit and the ancilla resonator as $\hat H_{\text{sa},\mathrm{eff}} = (\chi _{\text{sa},\mathrm{eff}} /2 ) \hat \sigma _{\text{z}} ^{(gf)} \resop ^\dag \resop $, where $\hat \sigma _{\text{z}} ^{(gf)} = \ket { 1_{\text{L}} }\!\bra{1_{\text{L}} } - \ket { 0_{\text{L}} }\!\bra{0_{\text{L}} }$.
Exact diagonalization of the driven Hamiltonian in the rotating frame defined by the drive frequency yields \( \chi_{\text{sa},\mathrm{eff}} \approx -1.22~\mathrm{kHz} \), in good agreement with the perturbative estimate. Denote the readout drive detuning with respect to the resonator mode as $\Delta_{\mathrm{dr}} \equiv \omega_{\mathrm{dr}} - \omega_{\text{r}}$, and assume that the readout pulse is longer than the ancilla resonator lifetime, $\kappa_a^{-1} \sim 530~\mathrm{ns}$, (which is typically satisfied in experiments considered in this work), the logical qubit AC Stark shift $\delta \omega  _{\mathrm{dr,eff}} $ and dephasing rate $\gamma _{\mathrm{dr,eff}}$ induced by the ancilla drive can be computed as~\cite{Gambetta2006,Clerk2007} 
\begin{align}
-i \delta \omega  _{\mathrm{dr,eff}} - \gamma _{\mathrm{dr,eff}}
\simeq - i \chi _{sa,\mathrm{eff}} \frac{|f _{\text{dr}}|^{2} }{\left( - \Delta _{\mathrm{dr}} + \frac{\chi _{\rm sa,\mathrm{eff}}}{2} - i \frac{\kappa _{a} }{2}
\right) \left( - \Delta _{\mathrm{dr}} - \frac{\chi _{\rm sa,\mathrm{eff}}}{2} + i \frac{\kappa _{a} }{2}
\right) }
. 
\end{align}
To calibrate the drive amplitude used during readout, we perform a $g-f$ Ramsey on the data transmon under a continuous-wave drive on the ancilla resonator. We measured around $70$~kHz Stark shift on the data transmon in Ramsey experiment over a $4\,\mu$s span. The drive frequency is chosen to be on resonance with the checking resonator frequency conditioned on the checking qubit being in $\ket{f}$ state. This leads to a drive detuning of $\Delta _{\mathrm{dr}} \equiv  \omega _{\mathrm{dr}} - \resfreq = \chi_{ca,\text{bias}} ^{(gf)}  - \chi_{ca,\text{bias}} /2 \approx 0.64~$MHz. In this case, as the drive detuning of magnitude is much larger than the dispersive shift of data qubit, we can approximately write 
\begin{align}
-i \delta \omega  _{\mathrm{dr,eff}} - \gamma _{\mathrm{dr,eff}}
= - i \chi _{\rm sa,\mathrm{eff}}
\frac{|f _{\text{dr}}|^{2} }{\Delta _{\mathrm{dr}} ^{2} + \left ( \frac{\kappa _{a} }{2} \right) ^{2} }
\left [ 1 - \frac{i  \frac{\chi _{\rm sa,\mathrm{eff}}}{2} \kappa _{a}  }{\Delta _{\mathrm{dr}} ^{2} + \left ( \frac{\kappa _{a} }{2} \right) ^{2}} \right ]
.
\end{align}
We see that if the averaged Stark shift is $\delta \omega  _{\mathrm{dr,eff}} \approx 70$~kHz, then the photon shot-noise dephasing during the same period should equal 
\begin{align}
\gamma _{\mathrm{dr,eff}} =  \frac{  \frac{\chi _{\rm sa,\mathrm{eff}}}{2} \kappa _{a}  }{\Delta _{\mathrm{dr}} ^{2} + \left ( \frac{\kappa _{a} }{2} \right) ^{2}} \delta \omega  _{\mathrm{dr,eff}}
\approx 23~\text{Hz}
.
\end{align}
This leads to a photon-shot-noise dephasing rate of $\approx 1~\text{rad}/(7~\text{ms})$. This is consistent with experimental calibration measurements, i.e., high-fidelity erasure syndrome extraction does not lead to an observable change in the data qubit dephasing time within experimental uncertainty. 

\section{Analysis of error channels of single-qubit logical gate} 

\begin{table}[!h]
\caption{Infidelity contribution for an $80$\,ns $\pi$ pulse for no decoherence, with decoherence (pure dephasing channel), with decoherence (bit-flip/ decay channel), with decoherence (both decay, dephasing channels) and with decoherence (both decay, dephasing channels), imperfect post selection (PS) on $\ket{e}$ state, considering false negative rate}
\label{tab:rb_budgetpi}
\begin{ruledtabular}
\begin{tabular}{lccccc}
Source & No Decoherence & Pure Dephasing &Bit-Flip/ Decay channel & Decoherence& Decoherence 
\\ 
  & & &   & & (imperfect PS) \\
\hline
\hline

$\ket{e}$ leakage
& $2.72\times10^{-4}$ 
&  $2.76\times10^{-4}$
&$1.66\times10^{-3}$
&$1.66\times10^{-3}$
&$1.88\times10^{-4}$\\

$\ket{h}$ leakage
& $0.02\times10^{-4}$ 
&$0.03\times10^{-4}$ 
&$0.02\times10^{-3}$
&$0.02\times10^{-3}$
&$0.18\times10^{-4}$\\

Pauli error
& $--$ 
&$0.26\times10^{-4}$ 
&$0.08\times10^{-3}$
&$0.10\times10^{-3}$
&$1.07\times10^{-4}$\\

Control error
& $0.02\times10^{-4}$ 
& $--$
& $--$
& $--$
& $--$\\
%Residual
%& $1.76\times10^{-6}$ 
%& $--$
%& $--$
%& $--$
%\\
\hline
Simulated gate error 
& $2.76\times10^{-4}$ 
& $3.05\times10^{-4}$ 
& $1.76\times10^{-3}$
& $1.78\times10^{-3}$
& $3.13\times10^{-4}$\\
\end{tabular}
\end{ruledtabular}
\end{table}

\begin{table}[!h]
\caption{Infidelity contribution for an $80$\,ns $\pi /2$ pulse for no decoherence, with decoherence (pure dephasing channel), with decoherence (bit-flip/ decay channel), with decoherence (both decay, dephasing channels) and with decoherence (both decay, dephasing channels), imperfect post selection on $\ket{e}$ state, considering false negative rate}
\label{tab:rb_budgetpi2}
\begin{ruledtabular}
\begin{tabular}{lccccc}
Source & No Decoherence & Pure Dephasing &Bit-Flip/ Decay channel & Decoherence& Decoherence 
\\ 
  & & &   & & (imperfect PS) \\
\hline
\hline

$\ket{e}$ leakage
& $2.44\times10^{-5}$ 
&  $2.62\times10^{-5}$
&$1.48\times10^{-3}$
&$1.48\times10^{-3}$
&$1.68\times10^{-4}$\\

$\ket{h}$ leakage
& $0.09\times10^{-5}$ 
&$0.13\times10^{-5}$ 
&$0.01\times10^{-3}$
&$0.01\times10^{-3}$
&$0.15\times10^{-4}$\\

Pauli error
& $--$ 
&$2.92\times10^{-5}$ 
&$0.05\times10^{-3}$
&$0.07\times10^{-3}$
&$0.75\times10^{-4}$\\

Control error
& $0.33\times10^{-5}$ 
& $--$
& $--$
& $--$
& $--$\\
%Residual
%& $1.76\times10^{-6}$ 
%& $--$
%& $--$
%& $--$
%\\
\hline
Simulated gate error 
& $2.86\times10^{-5}$ 
& $5.67\times10^{-5}$ 
& $1.54\times10^{-3}$
& $1.57\times10^{-3}$
& $2.58\times10^{-4}$\\
\end{tabular}

\end{ruledtabular}
\end{table}

We perform numerical simulations of the \(g\)–\(f\) erasure qubit Hamiltonian to better understand the error channels contributing to logical gate infidelity. Specifically, we simulate \(\pi\) and \(\pi/2\) gates using Gaussian pulses augmented with the DRAG formalism to suppress leakage into non-computational states. We assume the primary sources of infidelity are leakage to non-computational states (such as \(\ket{e}\) and \(\ket{h}\)), Pauli errors—including bit-flip errors—undetected decay from \(\ket{e}\) to \(\ket{g}\), and pure dephasing errors. We compute the average infidelity over the six cardinal states of the Bloch sphere for both the \(\pi\) pulse (reported in Table~\ref{tab:rb_budgetpi}) and the \(\pi/2\) pulse (reported in Table~\ref{tab:rb_budgetpi2}). We report the total gate infidelity caused by different noise channels. For each channel, we account for contributions from leakage, Pauli errors, and control errors. Simulation parameters include \(T_{\phi}^{gf} = 440\,\mu s\) as reported in the main text. For decay channels, we use experimental decay times \(T_1^{ge} = 52\,\mu s\) and \(T_1^{ef} = 26\,\mu s\). To incorporate the thermal excitation rate, we use an average thermal population ($P_{th} = 0.7\%$), which yields an excitation rate \(\gamma_{up} = \frac{P_{th}}{T_1^{ge}} = 1.37 \times 10^{-7}\).\\

\begin{figure}[!thtbp]
	\centering
	\includegraphics[width=12cm]{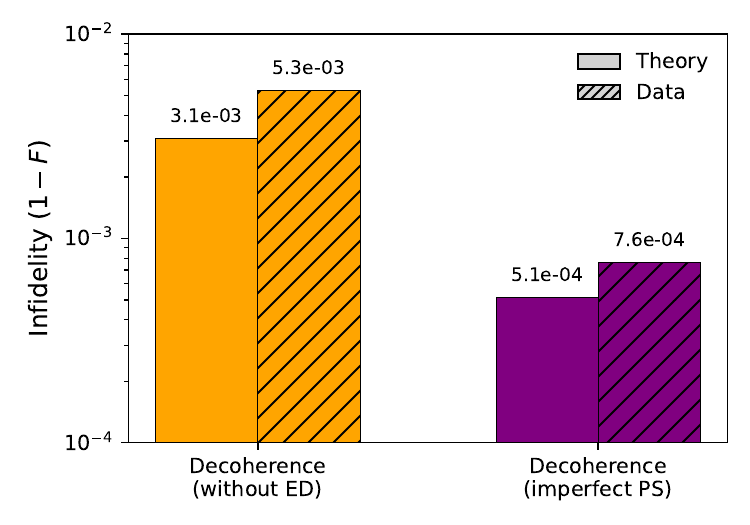}
	\caption{Bar graph comparing simulated and experimental average Clifford gate infidelities. The simulated infidelity without erasure detection is $2.95 \times 10^{-3}$, while the corresponding experimental Clifford gate infidelity without erasure checking is $5.3 \times 10^{-3}$ (left). On the right, the simulated infidelity including both dephasing and decay channels with imperfect post-selection is $4.91 \times 10^{-4}$, compared with the experimental Clifford gate infidelity with erasure checking of $7.6 \times 10^{-4}$.}\label{Fig:infid_compare}
\end{figure}

To closely model our experimental result without erasure detection and with erasure detection, we consider decoherence noise model with and without erasure detection. To model our experimental result without erasure detection, we use decoherence channel with both dephasing and decay channels. To model our experimental result with erasure detection we use decoherence noise model, with imperfect postselection on $\ket{e}$ state. To account for the imperfect postselection our simulation employs a false negative rate of $\epsilon_{fn}\approx0.17$ from our RB experiment, larger than the rates observed in logical lifetime experiments. And we consider $1/3$ of the false negative events could be detected in the next checking cycle thus is not harmful as explained in our analysis of bit-flip errors. Our 24 Clifford gates are composed of 45 physical gates, giving use $1.875$ physical gates per Clifford gate. Out of our 45 physical gates we have $8$ ($\pi$) gates, $1$ Identity gate and $36$ ($\pi/2$) gates. So our simulated error per Clifford gate in terms our our physical gate infidelity is given by: $ \frac{45}{24}\times(\frac{8}{45} \epsilon_{\pi} + \frac{1}{45} \epsilon_{I} + \frac{36}{45} \epsilon_{\pi/2}) $. We get a physical gate infidelity for Identity gate of $1.57\times10^{-3}$ without post selection and $2.33\times10^{-4}$ with post selection. Using this relation we report our simulated infidelity per Clifford gate in Fig.~\ref{Fig:infid_compare}, contrasted with our experimental results. The discrepancy between simulation and experimental results can be attributed to extra noise channels in the experiments, such as frequency, coherence fluctuations.

% In Fig.4, we show a Clifford gate infidelity of $5.3\times10^{-3}$ measured using RB without erasure checking gate. To model this result we use the Decoherence noise model with both dephasing and decay channels. Using the average gate infidelities of both $\pi$ and $\pi/2$ pulses, we extract simulated Clifford infidelity of $3.07\times10^{-3}$ as can be seen in Fig.\ref{Fig:infid_compare} (left column). For the case of interleaved RB with erasure checking, we get an average Clifford gate infidelity of $7.6\times10^{-4}$. To model this result we use Decoherence noise model with both dephasing and decay channels, with imperfect postselction on $\ket{e}$ state. The simulation employs a false negative rate of $\epsilon_{fn}\approx0.17$ from our RB experiment, larger than the rates observed in logical lifetime experiments. And we consider $1/3$ of the false negative events could be detected in the next checking cycle thus is not harmful similar to analysis of bit-flip error.
% Using these parameters we get an average simulated Clifford gate infidelity of $5.14\times10^{-4}$ as shown in Fig.\ref{Fig:infid_compare} (right column). The discrepancy of between theory and experimental infidelity can be attributed to extra noise channels in the experiments, such as frequency fluctuation of the flux tunable data qubit, erasure gate.\\

\section{Coherence stability of $g$--$f$ erasure qubit}
\begin{figure*}[!thtbp]
	\centering
	\includegraphics[width=18cm]{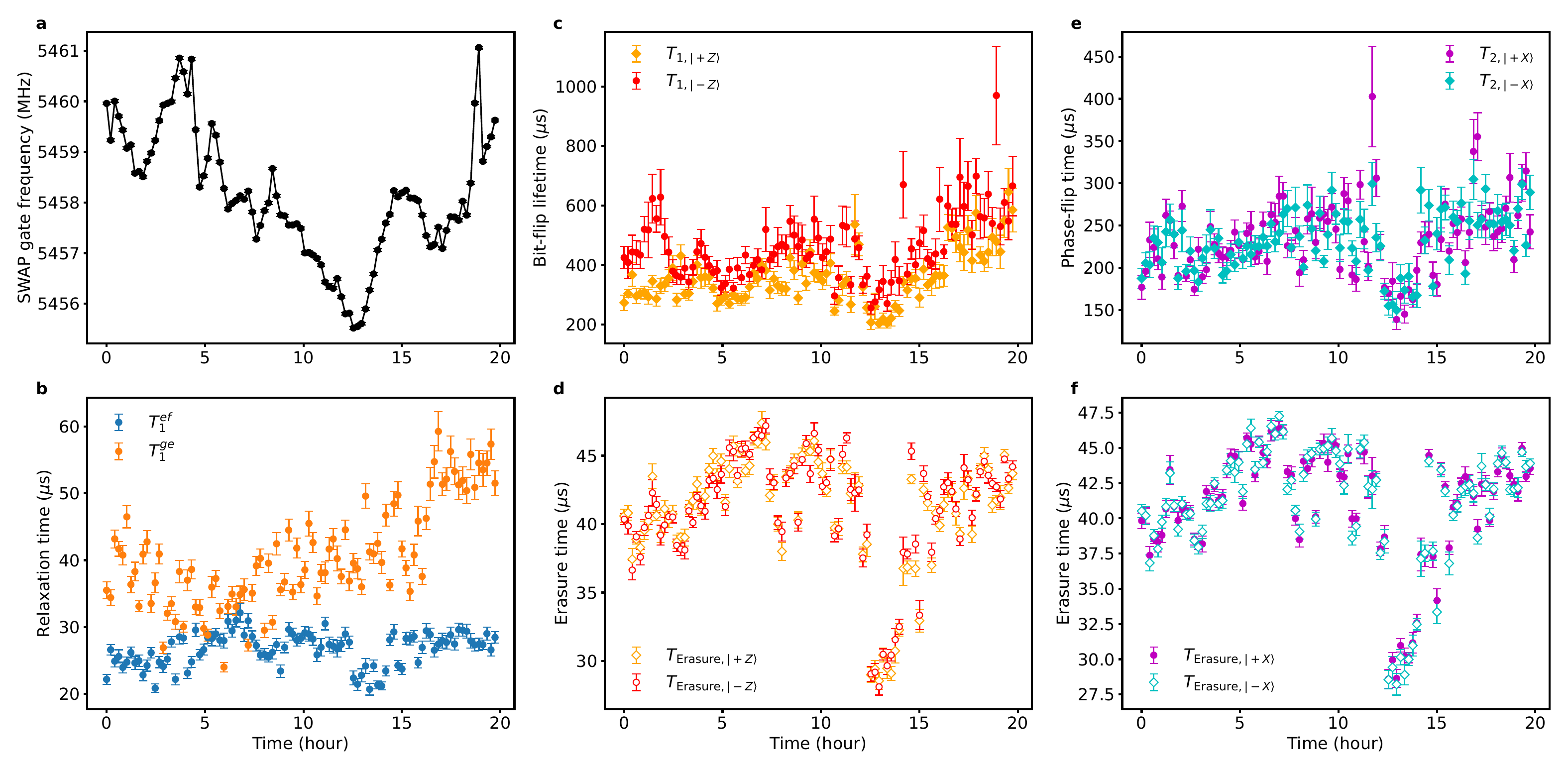}
	\caption{ Coherence stability.  \textbf{a}, The frequency of erasure SWAP gate (200\,ns) was fitted from spectroscopy of $|eg\rangle\leftrightarrow|0_{\rm L}f\rangle$ transition. \textbf{b}, Fluctuations of $T^{ge}_{1}$ and $T^{ef}_{1}$ decay. \textbf{c}, bit-flip time $T_{1,|\pm Z\rangle}$ and \textbf{e} phase-flip time $T_{1,|\pm X\rangle}$, with the erasure lifetime \textbf{d} and \textbf{f}, respectively.
    Each data point represents the average of two experimental runs with recalibration.}\label{Stability}
\end{figure*}

The repeated memory lifetime of $g$–$f$ erasure qubit was taken over the 20-hour measurement window, showing only slow, correlated drifts across all measured quantities. The syndrome frequency of erasure SWAP gate fluctuates gradually at the MHz level (see Fig.~\ref{Stability}a), which likely reflects flux noise, since the qubit operates several hundred MHz away from its zero-flux point. As shown in Fig.~\ref{Stability}b, relaxation times $T_1^{ge}$ and $T_1^{ef}$ exhibit small, gradual variations but remain within a stable range, with no sudden drops. Bit-flip and phase-flip lifetimes follow the similar slow trends, changing smoothly (see Fig.~\ref{Stability}c and Fig.~\ref{Stability}e). At approximately 13–15 hours, a small dip is observed across most measurements of the relaxation, erasure lifetime, bit-flip lifetime, and phase-flip times in Fig.~\ref{Stability}b–f, likely caused by transient interactions with a TLS defect coupled to the qubit.

\section{Cross-Resonance CNOT Gate }

The cross-resonance interaction serves as a key for implementing parity measurements between the data qubit and the ancilla~\cite{Blok2021}. Cross-resonance gate is realized by driving a data qubit (control) at the resonance frequency of an ancilla qubit, with an additional cancellation pulse on ancilla qubit (target)~\cite{sheldon2016procedure}, when the data qubit is in the state $\lvert 0_{\text{L}} \rangle$. Under these conditions, the system is described by the effective Hamiltonian  in a block-diagonal form $H_{\mathrm{CR}}/\hbar
= \frac{\Delta_{ac}}{2} \, (\lvert 0_{\text{L}} \rangle \langle 0_{\text{L}} \rvert-\lvert 1_{\text{L}} \rangle \langle 1_{\text{L}} \rvert) \otimes I
+ \frac{\Omega_{eff}}{2} \, \lvert 0_{\text{L}} \rangle \langle 0_{\text{L}} \rvert \otimes \sigma_{x}$,  where $I$ and $\sigma_{x}$ are Pauli operator on the ancilla qubit, \(\Omega_{eff}\) denote effective drive-dependent coupling strengths and $\Delta_{ac}$ represents an ac Stark shift on data qubit. To compensate for the ac Stark–shift–induced phase error, a virtual Z-phase gate can be applied to the data qubit.  
Note that off-resonant Rabi oscillations of the ancilla qubit can occur due to the CR drive when the data qubit is in the $|e\rangle$ state; these oscillations can be mittgated by appropriately selecting the drive amplitude, duration, and frequency, as discussed in Ref.~\cite{Blok2021}.

\end{document}